\documentclass[lettersize,journal]{IEEEtran}
\usepackage{amssymb}
\usepackage{amsmath,amsfonts}
\usepackage{algorithmic}
\usepackage{algorithm}
\usepackage{array}
\usepackage[caption=false,font=normalsize,labelfont=sf,textfont=sf]{subfig}
\usepackage{textcomp}
\usepackage{stfloats}
\usepackage{url}
\usepackage{verbatim}
\usepackage{graphicx}
\usepackage{cite}
\usepackage{booktabs}
\usepackage{multicol}
\usepackage{graphicx}
\usepackage[export]{adjustbox}
\usepackage{array}
\usepackage{subcaption}
\usepackage{xcolor}
\newcolumntype{P}[1]{>{\centering\arraybackslash}p{#1}}

\usepackage{helvet} 

\usepackage{setspace} 
\usepackage{placeins}

\usepackage{tikz}
\usetikzlibrary{shapes,arrows,positioning,calc}

\usepackage{afterpage}

\tikzset{
  block/.style = {rectangle, draw, fill=white, text centered, rounded corners, minimum height=2em, minimum width=1em}, 
  block2/.style = {rectangle, draw, fill=white, text centered, rounded corners, minimum height=2em, minimum width=1em},
  input/.style = {coordinate}, 
  output/.style = {coordinate},
  sum/.style = {draw, circle, inner sep=0pt, minimum size=0.5cm}, 
  sum2/.style = {draw, circle, inner sep=0pt, minimum size=0.5cm},
  arrow/.style = {->, thick, >=latex} 
}

\usepackage{hyperref}

\usepackage{bm}

\begin{document}

\bstctlcite{IEEEexample:BSTcontrol} 

\title{Airborne Particle Communication Through Time-Varying Diffusion-Advection Channels}

\author{Fatih~Merdan,
       Ozgur B.~Akan,~\IEEEmembership{Fellow,~IEEE}

\thanks{Fatih Merdan and O.B. Akan are with the Centre for neXt
Communications (CXC), Department of Electrical and Electronics Engineering, Koc¸ University, Istanbul 34450, Türkiye (e-mail:  fmerdan25@ku.edu.tr, akan@ku.edu.tr).}
\thanks{O. B. Akan is also with the Centre for neXt Communications (CXC), Department of Engineering, University of Cambridge, Cambridge CB3 0FA, United Kingdom (e-mail: oba21@cam.ac.uk).}}

\markboth{}%
{}

\maketitle

\begin{abstract}
Particle-based communication using diffusion and advection has emerged as an alternative signaling paradigm recently. While most existing studies assume constant flow conditions, real macro-scale environments such as atmospheric winds exhibit time-varying behavior. In this work, airborne particle communication under time-varying advection is modeled as a linear time-varying (LTV) channel, and a closed-form, time-dependent channel impulse response is derived using the method of moving frames. Based on this formulation, the channel is characterized through its power delay profile, leading to the definition of channel dispersion time as a physically meaningful measure of channel memory and a guideline for symbol duration selection. System-level simulations under directed, time-varying wind conditions show that waveform design is critical for performance, enabling multi-symbol modulation using a single particle type when dispersion is sufficiently controlled. To quantify waveform distortion and guide the design of orthogonal signaling waveforms, the Orthogonality Loss Ratio (OLR) is introduced as a structural metric. The results demonstrate that time-varying diffusion–advection channels can be systematically modeled and engineered using communication-theoretic tools, providing a realistic foundation for particle-based communication in complex flow environments.

\end{abstract}


\section{Introduction}

\IEEEPARstart{P}{ropagation} of energy is used to convey information in conventional communication systems. In contrast, unconventional communication paradigms rely on the propagation of mass to transfer information. Despite the fundamentally different physical mechanisms, both paradigms can be analyzed using the same mathematical framework provided by communication theory. Consequently, the concept of using mass as an information–carrying quantity has been primarily investigated within the literature of molecular communication (MC). MC proposes a communication paradigm in which molecules are used to encode, transmit, and receive information at both the micro- and macro-scale \cite{Receiver_Design_for_Molecular_Communication}. Until recently, most studies in MC have focused on the micro-scale domain \cite{A_Physical_EndtoEnd_Model_for_Molecular_Communication_in_Nanonetworks, Fundamentals_of_Molecular_Information_and_Communication_Science}. However, macro-scale molecular communication has started to attract increasing attention, particularly in scenarios where conventional electromagnetic communication is impractical or unreliable \cite{Macroscale_Molecular_Communication_in_IoTBased_Pipeline_Inspection_and_Monitoring_Applications_Preliminary_Experiment_and_Mathematical_Model}. Moreover, odor-based molecular communication (OMC), a specialized macro-scale MC paradigm in which information carriers are specific odor molecules, has been widely used to model inter- and intra-species communication in natural systems \cite{EndtoEnd_Mathematical_Modeling_of_Stress_Communication_Between_Plants}. Although MC is often motivated by biological contexts, the use of mass to transfer information is a more general concept. For this reason, the term particle is used throughout this paper instead of molecule. In the micro-scale, diffusion alone can be sufficient to support information transfer. In contrast, at the macro-scale, diffusion by itself is inadequate, and advective transport must be explicitly considered \cite{Modulation_Analysis_in_MacroMolecular_Communications}. The characterization of advection in diffusion–advection channels—particularly under realistic, time-varying flow conditions—remains an open problem in the context of particle-based communication.

Existing MC channel models can be broadly categorized into three main classes. The first class consists of constant-flow models, where advection is assumed to be steady and uniform, leading to analytically tractable channel descriptions \cite{Channel_Modeling_for_Diffusive_Molecular_CommunicationA_Tutorial_Review}. These models neglect environmental variability. The second class includes deterministic time-varying advection models, where the drift velocity varies over time but is assumed to be known and prescribed.  Recent works have developed analytical channel models under such conditions, capturing nonstationary transport effects induced by time-varying drift profiles \cite{Corrected_Inverse_Gaussian_First_Hitting_Time_Modeling_for_Molecular_Communication_Under_Time_Varying_Drift}. In parallel, stochastic formulations based on particle-level randomness, such as reaction–diffusion and Langevin-type models, have been investigated to capture microscopic fluctuations \cite{Stochastic_Reaction_and_Diffusion_Systems_in_Molecular_Communications_Recent_Results_and_Open_Problems}, \cite{An_Information_Theoretic_Framework_to_Analyze_Molecular_Communication_Systems_Based_on_Statistical_Mechanics}. The third class consists of simulation-based approaches, where realistic flow conditions—particularly turbulent environments—are modeled using computational fluid dynamics (CFD) or particle-based simulations \cite{A_Computational_Approach_for_the_Characterization_of_Airborne_Pathogen_Transmission_in_Turbulent_Molecular_Communication_Channels}, \cite{Efficient_Simulation_of_Macroscopic_Molecular_Communication}. While these approaches capture complex transport phenomena, they typically do not yield closed-form analytical channel models. Although deterministic time-varying models provide analytical tractability under prescribed drift profiles, and CFD-based approaches capture realistic turbulent behavior through numerical simulations, an analytical channel model that incorporates stochastic, time-varying advection remains largely unexplored. The proposed framework fills this gap by retaining analytical tractability and enabling statistical characterization across a broader range of stochastic advection conditions. In contrast to existing stochastic molecular communication models, which primarily account for randomness at the particle level via diffusion or stochastic perturbations, the proposed framework models the advection field itself as a stochastic process. This enables the characterization of particle communication channels under random time-varying flow conditions.

To address the limitations of existing channel models, it is essential to consider analytical techniques that enable tractable treatment of advection in diffusion–advection systems. A common approach is to apply a coordinate transformation, known as the method of moving frames, which simplifies the drift by shifting to a reference frame moving with the flow. This technique has been employed in MC under constant drift as well as in systems with deterministically time-varying transport mechanisms. Several studies have utilized such transformations in the context of electrophoretic molecular communication, where time-varying electric fields induce controlled drift profiles \cite{Electrophoretic_Molecular_Communication_With_Piecewise_Constant_Electric_Field, Electrophoretic_Molecular_Communication_with_TimeVarying_Electric_Fields, Molecular_Communications_Enhanced_by_TimeVarying_Electric_Field}. Similar formulations have also been adopted in classical drift–diffusion channel models and in scenarios involving moving propagation media \cite{Molecular_Communication_Using_Brownian_Motion_With_Drift, NanoMachine_Molecular_Communication_over_a_Moving_Propagation_Medium, Diffusive_Molecular_Communication_with_Disruptive_Flows}. Extending this framework to stochastic settings introduces a fundamentally different modeling paradigm, in which the transformed dynamics inherit randomness from the underlying flow field, and the resulting channel impulse response becomes a stochastic quantity.

From a communication-theoretic perspective, modulation in MC has traditionally been realized along specific physical dimensions, such as concentration, molecule type, and release timing \cite{Modulation_Techniques_for_Communication_via_Diffusion_in_Nanonetworks, A_Survey_on_Modulation_Techniques_in_Molecular_Communication_via_Diffusion, On_the_Characterization_of_Binary_ConcentrationEncoded_Molecular_Communication_in_Nanonetworks, Spatiotemporal_Distribution_and_Modulation_Schemes_for_ConcentrationEncoded_MediumtoLong_Range_Molecular_Communication}. Higher-order signaling schemes, including $M$-ary concentration shift keying (CSK), have been proposed to increase data rates \cite{An_MAry_ConcentrationShift_Keying_With_Common_Detection_Thresholds_for_Multitransmitter_Molecular_Communication}. Although these approaches enable richer signaling strategies, the selection of key communication parameters such as transmitter–receiver distance and symbol duration remains largely heuristic. In particular, the dispersion of particle signals induced by diffusion–advection channels is not characterized by a unified analytical metric that directly links physical transport properties to communication performance. In this work, we address this limitation by introducing channel dispersion time, which provides an analytical measure of signal spread in diffusion–advection channels. This metric enables systematic determination of communication parameters, including feasible transmitter–receiver separation and symbol duration, under given environmental conditions. Furthermore, the proposed framework enables a more systematic utilization of the signaling space by using orthogonal pulses, similar to those used in classical digital communication theory. In particular, it allows for the construction of multidimensional and higher-cardinality modulation schemes using a single particle type by designing signaling waveforms that are orthogonal at the receiver, analogous to orthogonal pulse shaping in classical digital communication. To quantify the extent to which this orthogonality is preserved after propagation through diffusion–advection channels and practical receiver dynamics, we introduce the Orthogonality Loss Ratio (OLR), which serves as a structural metric for evaluating pulse separability. This contrasts with conventional approaches in MC, where increased modulation dimensionality is typically achieved through multiple molecule types or additional spatial resources, rather than through pulse orthogonality. Although the primary focus of this work is on macro-scale particle communication, the developed framework is general and can also be applied to micro-scale systems. However, practical implementation in nanoscale environments requires careful consideration of receiver complexity and computational constraints, which may limit the direct applicability of advanced signal processing techniques.

The remainder of this paper is organized as follows. Section II presents the analytical solution to the time-varying diffusion–advection channel. Section III discusses the implications of the analytical model and characterizes the channel dispersion time. In Section IV, a representative communication system is analyzed under directed wind conditions. Section V introduces the OLR and analyzes particle pulse design under both channel and practical receiver dynamics. Concluding remarks are given in Section VI.

\section{time-varying diffusion-advection Channel}
\subsection{Time-varying Channel Impulse Response}
The behaviour of particles under diffusion and advection is described with a partial differential equation as
\begin{equation}
\label{eq:pde_c}
\begin{split}
\frac{\partial c}{\partial t}(\mathbf{x}, t)
&= D \nabla^{2} c(\mathbf{x}, t)
   - \mathbf{v}(t) \cdot \nabla c(\mathbf{x}, t)
   + S(\mathbf{x}, t),
\end{split}
\end{equation}
\noindent where $c(\mathbf{x},t)$ denotes the concentration of the particle at position \(\mathbf{x}\) and time \(t\), $D$ is the constant diffusion coefficient, $\mathbf{v}(t)$ is the wind velocity vector, $\nabla c(\mathbf{x},t)$ and $\nabla^{2} c(\mathbf{x},t)$ represent the spatial gradient and Laplacian of the concentration, respectively, and $S(\mathbf{x},t)$ denotes the external source term. To find an analytical expression for a time-varying diffusion-advection channel, the advection term is assumed to be constant in all space but varying in time. The position vector \(\mathbf{x}\) is defined as \(\mathbf{x} = (\mathbf{x}_{\parallel}, x_3) \in \mathbb{R}^{2} \times (0, \infty)\). It is assumed that \(x_3=0\) is a perfectly absorbing boundary i.e., \(c(\mathbf{x}_{\parallel}, 0, t)= 0\) and \( c(\mathbf{x}, 0) = c_{0}(\mathbf{x})\) is the initial particle concentration before transmission. Similarly, the wind velocity vector is defined as \(\mathbf{v}(t)= \big(\mathbf{v}_{\parallel}(t), 0\big)\) so that the \(x_3\) component of the wind is \(0\). This means that the gradient term in (\ref{eq:pde_c}) can be rewritten as
\begin{equation}
\label{eq:drift_term_parallel}
\begin{split}
- \mathbf{v}(t) \cdot \nabla c(\mathbf{x}, t)
&= - \mathbf{v}_{\parallel}(t)
   \cdot \nabla_{\mathbf{x}_{\parallel}} c(\mathbf{x}, t).
\end{split}
\end{equation}
Defining \( \mathbf{r}_{\parallel}(t) \overset{\triangle}{=} \int_{0}^{t} \mathbf{v}_{\parallel}(\xi)\, d\xi \), the moving frame variables are introduced as \(\mathbf{y}_{\parallel}= \mathbf{x}_{\parallel} - \mathbf{r}_{\parallel}(t)\), \(y_{3}=x_{3}\) and \(\tau= t\). The transformed concentration field is defined using these variables as \(u(\mathbf{y}, t)
= c\big(\mathbf{y}_{\parallel} + \mathbf{r}_{\parallel}(t),\, y_{3},\, t\big)\). Defining \(\mathbf{y}= \big(\mathbf{y}_{\parallel}, y_{3}\big)= \big(\mathbf{x}_{\parallel} - \mathbf{r}_{\parallel}(t),\, x_{3}\big)\), the chain rules for dimensions \(x_1\) and \(x_2\) can be written as
\begin{equation}
\label{eq:chain_rule_ct}
\begin{split}
\frac{\partial}{\partial t} c(x,t)
= \frac{\partial}{\partial t}\,
u\bigl(y(x,t),\, \tau(t)\bigr)
= u_{\tau}(y,\tau)\, \frac{\partial \tau}{\partial t}
  + u_{y}(y,\tau)\, \frac{\partial y}{\partial \tau}.
\end{split}
\end{equation}
Using the Leibniz rule, \(\frac{\partial y}{\partial \tau}= -\frac{d r}{d\tau}(\tau) = -v(\tau)\) is obtained. Therefore, (\ref{eq:chain_rule_ct}) becomes
\begin{equation}
\label{eq:ct_final_chain}
\begin{split}
\frac{\partial}{\partial t} c(x,t)
&= u_{\tau}(y,\tau)
   - v(\tau)\, u_{y}(y,\tau).
\end{split}
\end{equation}
Moreover, since \(\frac{\partial y}{\partial x} = 1\)  and $\frac{\partial \tau}{\partial x} = 0$, \( \frac{\partial}{\partial x} c(x,t) = u_{y}(y,\tau)\) and \(\frac{\partial^{2}}{\partial x^{2}} c(x,t) = u_{yy}(y,\tau)\) are obtained. These relations extend componentwise to the vector case, giving
\begin{equation}
\label{eq:vector_relations_combined}
\begin{split}
\frac{\partial}{\partial \tau} c(\mathbf{x}_{\parallel},\tau)
&= \frac{\partial}{\partial \tau} u(\mathbf{y}_{\parallel},\tau)
   - \mathbf{v}_{\parallel}(\tau)\cdot
     \nabla_{\mathbf{y}_{\parallel}} u(\mathbf{y}_{\parallel},\tau), \\[4pt]
\nabla_{\mathbf{x}_{\parallel}} c(\mathbf{x}_{\parallel},\tau)
&= \nabla_{\mathbf{y}_{\parallel}} u(\mathbf{y}_{\parallel},\tau),
\quad
\nabla^{2}_{\mathbf{x}_{\parallel}} c(\mathbf{x}_{\parallel},\tau)
= \nabla^{2}_{\mathbf{y}_{\parallel}} u(\mathbf{y}_{\parallel},\tau).
\end{split}
\end{equation}
For the $x_{3}$ dimension one can obtain
\begin{equation}
\label{eq:u_y3_and_derivatives}
\begin{split}
u(y_{3},\tau)
&= c(x_{3},t), \\[4pt]
\frac{\partial}{\partial t} c(x,t)
&= \frac{\partial u}{\partial \tau}\,
   \frac{\partial \tau}{\partial t}
 + \frac{\partial u}{\partial y}\,
   \frac{\partial y}{\partial t}
 = \frac{\partial u}{\partial \tau}(y,\tau), \\[4pt]
\frac{\partial}{\partial x} c(x,t)
&= \frac{\partial u}{\partial y}(y,\tau), 
\,\,\,\,\frac{\partial^{2}}{\partial x^{2}} c(x,t)
= \frac{\partial^{2} u}{\partial y^{2}}(y,\tau).
\end{split}
\end{equation}
Thus, using $t$ instead of $\tau$ for simplicity with \(\mathbf{y}=(\,\mathbf{y}_{\parallel},\; y_{3}\,)=(\,\mathbf{x}_{\parallel} - \mathbf{r}_{\parallel}(t),\; x_{3}\,)\) and \(\frac{\partial \mathbf{y}}{\partial t}
= (-\frac{\partial \mathbf{r}_{\parallel}}{\partial t},\; 0)
 = (-\mathbf{v}_{\parallel}(t),\; 0)\) one can obtain 
\begin{equation}
\label{eq:ct_vector_final}
\begin{split}
\frac{\partial}{\partial t} c(\mathbf{x},t)
&= \frac{\partial}{\partial t} u(\mathbf{y},t)
   - \mathbf{v}_{\parallel}(t)
     \cdot \nabla_{\mathbf{y}_{\parallel}} u(\mathbf{y},t) .
\end{split}
\end{equation}
Inserting (\ref{eq:vector_relations_combined}),(\ref{eq:u_y3_and_derivatives}) and (\ref{eq:ct_vector_final}) into (\ref{eq:pde_c}) yields the new PDE as
\begin{equation}
\label{eq:pde_u}
\begin{split}
\frac{\partial}{\partial t} u(\mathbf{y}, t)
&= D \nabla_{\mathbf{y}}^{2} u(\mathbf{y}, t)
  + S\!\big(\mathbf{y}_{\parallel} + \mathbf{r}_{\parallel}(t),\, y_{3},\, t\big),
\end{split}
\end{equation}
\noindent with the initial and boundary conditions \( u(\mathbf{y}, 0) = c_{0}(\mathbf{y}), u(\mathbf{y}_{\parallel}, 0, t) = 0 \), respectively. This is a heat equation, and its free space kernel is found as \cite{Partial_Differential_Equations}
\begin{equation}
\label{eq:heat_kernel}
\Phi(\mathbf{y}, t) =
\begin{cases}
\dfrac{1}{(4 \pi D t)^{3/2}}
\exp\!\left(-\dfrac{\|\mathbf{y}\|^{2}}{4 D t}\right),
& \mathbf{y} \in \mathbb{R}^{3},\; t > 0, \\[10pt]
0, & t < 0 .
\end{cases}
\end{equation}
Using the method of images, the solution for (\ref{eq:pde_u}) is written as
\begin{equation}
\label{eq:u_solution_compact}
\begin{split}
u(\mathbf{y}, t)
&= \int_{\mathbb{R}^{3}}
\Phi_{D}(\mathbf{y},\mathbf{z};t)\,
c_{0}(\mathbf{z})\, d\mathbf{z} \\[4pt]
&\quad
+ \int_{0}^{t} \int_{\mathbb{R}^{3}}
\Phi_{D}(\mathbf{y},\mathbf{z};t - s)\,
S\!\big(\mathbf{z}_{\parallel} + \mathbf{r}_{\parallel}(s),\, z_{3},\, s\big)\,
d\mathbf{z}\, ds,
\end{split}
\end{equation}
\noindent where
\begin{equation}
\label{eq:Phi_D_def}
\begin{split}
\Phi_{D}(\mathbf{y},\mathbf{z};t)
&\overset{\triangle}{=} \Phi\!\big((\mathbf{y}_{\parallel} - \mathbf{z}_{\parallel},\, y_{3} - z_{3}),\, t\big) \\
&\quad
   - \Phi\!\big((\mathbf{y}_{\parallel} - \mathbf{z}_{\parallel},\, y_{3} + z_{3}),\, t\big) .
\end{split}
\end{equation}
The solution in the original coordinates can be written as
\begin{equation}
\label{eq:c_solution_general_phiD}
\textcolor{black}{
\begin{split}
c(\mathbf{x}, t)
&= \int_{\mathbb{R}^{3}}
\Phi_{D}\big( (\mathbf{x}_{\parallel} - \mathbf{r}_{\parallel}(t),\, x_{3}),\,
              \mathbf{z};\, t \big)\,
c_{0}(\mathbf{z})\, d\mathbf{z} \\[4pt]
&\quad
+ \int_{0}^{t} \int_{\mathbb{R}^{3}}
\Phi_{D}\big( (\mathbf{x}_{\parallel} - \mathbf{r}_{\parallel}(t)
               + \mathbf{r}_{\parallel}(s),\, x_{3}),\,
               \mathbf{z};\, t - s \big) \\
&\qquad\qquad\qquad \times
S\!\big(\mathbf{z},\, s\big)\,
d\mathbf{z}\, ds,
\end{split}
}
\end{equation}
\noindent where the \(\mathbf{r}_{\parallel}(s)\) terms inside \(\Phi_{D}\) originate from the motion of the source relative to the moving coordinate system. Assuming a point source at position \(\mathbf{z}_{0}\) i.e., \(S(\mathbf{z}, s)
= q(s)\,\delta(\mathbf{z} - \mathbf{z}_{0})\) and zero initial concentration for the particle, i.e., \(c_{0}(\mathbf{z}) = 0\), the result simplifies as
\begin{equation}
\label{eq:c_solution_point_source_phiD}
\begin{split}
c(\mathbf{x}, t)
&= \int_{0}^{t}
\Phi_{D}\big(
(\mathbf{x}_{\parallel} - \mathbf{r}_{\parallel}(t)
 + \mathbf{r}_{\parallel}(s),\, x_{3}),\,
\mathbf{z}_{0};\, t - s
\big)\,
q(s)\, ds .
\end{split}
\end{equation}
From (\ref{eq:c_solution_point_source_phiD}), for a fixed receiver position, the time-varying channel impulse response can be written as shown in (\ref{eq:LTV_impulse}). From (\ref{eq:LTV_impulse}), it is evident that the time-dependent nature of the diffusion-advection channel is due to variations in the wind with time. Under the assumption of constant wind, (\ref{eq:LTV_impulse}) yields the impulse response of the Linear Time-Invariant (LTI) channel. In that case, (\ref{eq:c_solution_point_source_phiD}) becomes a convolution of \(h(t)\) with the input to the channel \(q(t)\). This assumption may be fine for a very controlled environment, but, in general, it is not possible to control the wind flow with high accuracy. The Navier-Stokes equations govern the motion of viscous fluids and, in principle, can be used to determine the airflow velocity. However, their chaotic dynamics motivate modeling wind as a random process. Wind speed is commonly represented by a Weibull distribution, although no single distribution adequately describes every wind regime \cite{A_Review_of_Wind_Speed_Probability_Distributions_Used_in_Wind_Energy_Analysis_Case_Studies_in_the_Canary_Islands}. For the closed-form statistical analysis, unless otherwise stated, the two horizontal wind components are modeled as independent wide-sense stationary (WSS) white Gaussian processes with mean \(\mu\) and covariance \(q_v^2\delta(t-t')\). Section~II-B first formulates the channel autocorrelation for general independent wind components and invokes Gaussianity only for closed-form evaluation. Section~III-A relaxes only the white-wind assumption to examine colored WSS Gaussian wind.

\begin{figure*}[!t]
\small
\begin{equation}
\label{eq:LTV_impulse}
h(\tau,t) = \frac{1}{(4\pi D \tau)^{3/2}}
\Bigg[
\exp\!\left(
-\frac{
\left\|
\mathbf{x}_{\parallel}
- \mathbf{z}_{0\parallel}
- \displaystyle\int_{t-\tau}^{t} \mathbf{v}_{\parallel}(\xi)\, d\xi
\right\|^{2}
+ (x_{3} - z_{03})^{2}
}{4D\tau}
\right)
 -
\exp\!\left(
-\frac{
\left\|
\mathbf{x}_{\parallel}
- \mathbf{z}_{0\parallel}
- \displaystyle\int_{t-\tau}^{t} \mathbf{v}_{\parallel}(\xi)\, d\xi
\right\|^{2}
+ (x_{3} + z_{03})^{2}
}{4D\tau}
\right)
\Bigg].
\end{equation}
\end{figure*}

\subsection{Statistical characterization of the time-varying Impulse Response}

The derivation of the mean and autocorrelation of (\ref{eq:LTV_impulse}) requires a preliminary examination of the exponential components, as they contain the stochastic contributions of the wind. Upon expanding the norms, one obtains
\begin{equation}
\label{eq:expanded_xyz_resize}
\resizebox{0.90\linewidth}{!}{$
\begin{aligned}
\left( x_{1} - z_{01}
        - \int_{t-\tau}^{t} v_{1}(\lambda)\, d\lambda
\right)^{2}
&\;+\;
\left( x_{2} - z_{02}
        - \int_{t-\tau}^{t} v_{2}(\lambda)\, d\lambda
\right)^{2}
\\[6pt]
&\;+\;
(x_{3} - z_{03})^{2},
\end{aligned}
$}
\end{equation}
\noindent where \(v_1(\lambda)\) and \(v_2(\lambda)\) are i.i.d. wind processes with mean \(\mu\). Defining \(V_i(\tau,t)=\int_{t-\tau}^{t}v_i(\lambda)\,d\lambda\), its mean and variance are \(\mathbb{E}[V_i(\tau,t)]=\mu\tau\) and \(\operatorname{Var}(V_i(\tau,t))=\sigma_X^2(\tau,t)\), respectively, where
\begin{equation}
\label{eq:sigmaX_definition}
\begin{split}
\sigma^{2}_{X}(\tau,t)
&\overset{\triangle}{=} \int_{t-\tau}^{t} \int_{t-\tau}^{t}
    \mathrm{Cov}\!\big( v_{i}(\lambda_{1}),\, v_{i}(\lambda_{2}) \big)\,
    d\lambda_{1}\, d\lambda_{2},
\end{split}
\end{equation}
\noindent and \(\mathrm{Cov}\!\big( v_{i}(\lambda_{1}),\, v_{i}(\lambda_{2}) \big)\) is the covariance function of \(v_i\). To represent the random parts in (\ref{eq:expanded_xyz_resize}), \( X_i(t,\tau) \overset{\triangle}{=} x_i - z_{0i} - V_i \) are defined for \(i = 1, 2\). Accordingly, \(\mathbb{E}[X_i(\tau,t)]=x_i-z_{0i}-\mu\tau\) and \(\operatorname{Var}(X_i(\tau,t))=\sigma_X^2(\tau,t)\). Since \(v_1\) and \(v_2\) are independent, \(X_1\) and \(X_2\) are also independent. The autocorrelation of \(X_i\) is \(R_{X_i}(\tau_1,\tau_2;\,t_1,t_2)=(x_i-z_{0i}-\mu\tau_1)(x_i-z_{0i}-\mu\tau_2)+L(\tau_1,\tau_2;\,t_1,t_2),\)where
\begin{equation}
\label{eq:L_definition}
\begin{split}
L(\tau_{1},\tau_{2};\, t_{1},t_{2})
&\overset{\triangle}{=} \int_{t_{1}-\tau_{1}}^{t_{1}}
    \int_{t_{2}-\tau_{2}}^{t_{2}}
    \mathrm{Cov}\!\big( v_{1}(\lambda_{1}),\, v_{1}(\lambda_{2}) \big)\,
    d\lambda_{1}\, d\lambda_{2}.
\end{split}
\end{equation}
The time-varying impulse response in (\ref{eq:LTV_impulse}) can be written as \(h(\tau,t)
= \beta(\tau)\,
   \exp\!(
      -\alpha(\tau)\,
      [ X_{1}^{2}(\tau,t) + X_{2}^{2}(\tau,t) ]
   )\), where
\begin{equation}
\label{eq:h_beta_alpha_expanded}
\begin{split}
\beta(\tau)
= \frac{1}{(4\pi D\tau)^{3/2}}
   &\left(
      e^{-\frac{(x_{3}-z_{03})^{2}}{4D\tau}}
      - e^{-\frac{(x_{3}+z_{03})^{2}}{4D\tau}}
   \right),
\\[10pt]
\alpha(\tau)
&= \frac{1}{4D\tau},
\end{split}
\end{equation}
\noindent and the autocorrelation \(R_h(\tau_1,\tau_2;t_1,t_2)\) of \(h(\tau,t)\) can be expressed as shown in (\ref{eq:Rh_split_resize}). The autocorrelation form in (\ref{eq:Rh_split_resize}) is valid for arbitrary wind distributions, provided that the two horizontal wind components are independent. For Gaussian wind, these expectations can be evaluated analytically, yielding a closed-form autocorrelation expression. For $\mathbf{X} \sim \mathcal{N}_{p}(\boldsymbol{\mu},\boldsymbol{\Sigma})$ with $\boldsymbol{\Sigma} > 0$, and
$\mathbf{Q} = \mathbf{X}^{\top}\mathbf{A}\mathbf{X}$ where $\mathbf{A} = \mathbf{A}^{\top}$,
the moment generating function of $\mathbf{Q}$ is given as \cite{Quadratic_Forms_in_Random_Variables_Theory_and_Applications}
\begin{equation}
\label{eq:MGF_quadratic_form}
\resizebox{0.96\linewidth}{!}{$
\begin{aligned}
M_{\mathbf{Q}}(t)
&= \mathbb{E}\!\left[e^{t\mathbf{Q}}\right]
= \left|\,\mathbf{I}
      - 2t\,\boldsymbol{R}
  \,\right|^{-1/2}
\\[6pt]
&\quad\times
\exp\!\left(
t\,\boldsymbol{\mu}^{\top}\boldsymbol{\Sigma}^{-1/2}
      \boldsymbol{R}
      \bigl(\mathbf{I}
             - 2t\,\boldsymbol{R}
      \bigr)^{-1}
      \boldsymbol{\Sigma}^{-1/2}\boldsymbol{\mu}
\right),
\end{aligned}
$}
\end{equation}
\noindent where \( \boldsymbol{R} = \bigl(\boldsymbol{\Sigma}^{1/2}\mathbf{A}\boldsymbol{\Sigma}^{1/2}\bigr) \). Using \(p = 2\), inserting \(t = \frac{-1}{4D}\) and the following
\begin{equation}
\label{eq:A_mu_Sigma_definitions}
\begin{aligned}
\mathbf{X_i}
&=
\begin{bmatrix}
X_i(\tau_{1}, t_{1}) \\[4pt]
X_i(\tau_{2}, t_{2})
\end{bmatrix},
\quad
\boldsymbol{\mu_i}
=
\begin{bmatrix}
x_i - z_{0i} - \mu \tau_1 \\[4pt]
x_i - z_{0i} - \mu \tau_2
\end{bmatrix},
\\[12pt]
\boldsymbol{\Sigma}
&=
\begin{bmatrix}
\sigma^{2}_{X}(\tau_{1}, t_{1})
&
L(\tau_{1}, \tau_{2}; t_{1}, t_{2}) \\[6pt]
L(\tau_{1}, \tau_{2}; t_{1}, t_{2})
&
\sigma^{2}_{X}(\tau_{2}, t_{2})
\end{bmatrix},
\mathbf{A}
=
\begin{bmatrix}
\dfrac{1}{\tau_{1}} & 0 \\[4pt]
0 & \dfrac{1}{\tau_{2}}
\end{bmatrix},
\end{aligned}
\end{equation}
\noindent to (\ref{eq:MGF_quadratic_form}) we obtain an expression for \(R_{h}(\tau_{1},\tau_{2};\, t_{1},t_{2})\). However, when \(\tau_1 = \tau_2\) and \(t_1 = t_2\), \(\boldsymbol{\Sigma}\) becomes singular. The moment generating function of
\( \mathbf{Q} = \mathbf{X}^{\top}\mathbf{A}\mathbf{X} \),
where \(\mathbf{A} = \mathbf{A}^{\top}\), and the
\(p \times 1\) vector \(\mathbf{X}\) is normal with
\(\mathbb{E}[\mathbf{X}] = \boldsymbol{\mu}\) and
\(\mathrm{Cov}(\mathbf{X}) = \boldsymbol{\Sigma} = \mathbf{B}\mathbf{B}^{\top}\),
with \(\mathbf{B}\) being a \(p \times r\) matrix of rank \(r < p\), is given by \cite{Quadratic_Forms_in_Random_Variables_Theory_and_Applications}
\begin{equation}
\label{eq:log_MQ_split}
\begin{aligned}
&\ln M_{\mathbf{Q}}(t)
=
-\frac{1}{2}
\sum_{j=1}^{r}
\ln\!\bigl(1 - 2t\lambda_{j}\bigr)
+\;\alpha t
\;+\;
2t^{2}
\sum_{j=1}^{r}
\frac{b_{j}^{2}}{1 - 2t\lambda_{j}},
\end{aligned}
\end{equation}
\noindent where \(\lambda_{1},\ldots,\lambda_{r}\) are the eigenvalues of
\(\mathbf{B}^{\top}\mathbf{A}\mathbf{B}\),
with \(\mathbf{B}^{\top}\mathbf{A}\mathbf{B} \neq \mathbf{0}\), \(\alpha = \boldsymbol{\mu}^{\top}\mathbf{A}\boldsymbol{\mu} \), \(\mathbf{b}^{\top} = \mathbf{P}^{\top}\mathbf{B}^{\top}\mathbf{A}\boldsymbol{\mu} \) and \(\mathbf{P}\mathbf{P}^{\top} = \mathbf{I}\). Using \(p = 2\), \(r = 1\), the same \(\mathbf{A}\) and \(\boldsymbol{\mu}\) with \(\tau_1 = \tau_2\) from (\ref{eq:A_mu_Sigma_definitions}), inserting \(t = \frac{-1}{4D}\) and $\mathbf{B} = [\,\sigma_X\;\;\sigma_X\,]^{\top}$ to (\ref{eq:log_MQ_split}), we obtain the expression for \(R_{h}(\tau_{1},\tau_{2};\, t_{1},t_{2})\) where \(\tau_1 = \tau_2\) and \(t_1 = t_2\). A complete analytical expression for \(R_{h}(\tau_{1},\tau_{2};\, t_{1},t_{2})\) can be obtained with these two cases handled separately. Under the same Gaussian-wind assumption, the mean channel response can also be evaluated in closed form, yielding (\ref{eq:Eh_split}).

\afterpage{
\begin{figure*}[!t]
\begin{align}
R_{h}(\tau_{1},\tau_{2}; t_{1},t_{2})
&= \beta(\tau_{1})\,\beta(\tau_{2})\,
\mathbb{E}\!\left[
e^{-\alpha(\tau_{1})X_{1}^{2}(\tau_{1},t_{1})
   -\alpha(\tau_{2})X_{1}^{2}(\tau_{2},t_{2})}
\right]
\mathbb{E}\!\left[
e^{-\alpha(\tau_{1})X_{2}^{2}(\tau_{1},t_{1})
   -\alpha(\tau_{2})X_{2}^{2}(\tau_{2},t_{2})}
\right].
\label{eq:Rh_split_resize}
\\[8pt]
\mathbb{E}[h(\tau,t)]
&= \frac{\beta(\tau)}{1 + 2\alpha(\tau)\,(\sigma_X(\tau,t))^{2}}
\exp\!\left(
-\frac{
\alpha(\tau)\Big[(x_{1} - z_{01} - \mu\tau)^{2}
                + (x_{2} - z_{02} - \mu\tau)^{2}\Big]
}{
1 + 2\alpha(\tau)\,(\sigma_X(\tau,t))^{2}
}
\right).
\label{eq:Eh_split}
\end{align}
\end{figure*}
}

\subsection{Limitations and Physical Validity of Modeling Assumptions}

The analytical framework developed in this work relies on the assumption that the wind velocity is spatially uniform, i.e., $\mathbf{v}(\mathbf{x},t) \approx \mathbf{v}(t)$. Although this simplification enables a tractable derivation of the time-varying impulse response, its physical validity depends on the spatial coherence properties of turbulent flows. In atmospheric boundary layer turbulence, velocity fields exhibit a finite spatial correlation characterized by the \emph{integral length scale}, which characterizes the distance over
which the fluctuating velocity field is correlated \cite{Turbulent_Flows}. The integral length scale typically varies from 10 to 500 $m$ \cite{Atmospheric_Boundary_Layer_Flows_Their_Structure_and_Measurement, Investigations_of_Correlation_and_Coherence_in_Turbulence_from_a_LargeEddy_Simulation}. When the separation between the transmitter and receiver is sufficiently smaller than this characteristic scale, the spatial variation of the velocity field across the transmitter–receiver separation can be neglected, and the spatially uniform wind assumption becomes valid. Therefore, the proposed model is most appropriate for scenarios in which the transmitter–receiver separation is small relative to the dominant turbulence scales of the environment.

In addition to the spatial uniformity assumption, the wind velocity is modeled as a WSS white Gaussian process except for the wind covariance analysis in \autoref{Wind_channel_autocov_pairs} and \autoref{R_h_vs_tau2_different_time_seperations}. This assumption is adopted to enable a complete analytical characterization of the channel statistics. However, turbulent flows in the atmospheric boundary layer are not purely random fluctuations; instead, they are composed of coherent structures spanning a range of spatial and temporal scales. These structures, often referred to as eddies, introduce organized transport mechanisms and persistent flow patterns that are not fully captured by a Gaussian model. As a result, the Gaussian wind model should be interpreted as a simplified representation of the flow, while neglecting the detailed structure of turbulent motion. This simplification may smooth out localized or bursty transport effects associated with coherent structures, which can contribute to longer effective channel memory in practical scenarios. Nevertheless, the Gaussian model retains the dominant transport mechanisms through its mean and covariance structure, which govern the directed motion and dispersion of particles, and therefore captures the primary factors shaping the channel impulse response. Hence, the proposed model provides a physically meaningful and analytically tractable baseline for analyzing time-varying diffusion-advection channels. Incorporating non-Gaussian turbulence models or spatially varying velocity fields constitutes an important direction for future work.

\section{Statistical Analysis and Channel Dispersion Characterization}
\subsection{Effect of the Wind Covariance on the Channel Autocorrelation}
\begin{figure*}[!t]
  \centering
  \subfloat[]{
    \includegraphics[width=0.185\linewidth]{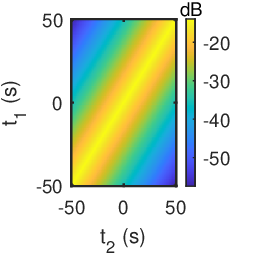} 
    \label{fig:WindCov2D_dB_WSSexp_Tv10.0_mu0.50_sig0.20}}
    \hfill
  \subfloat[]{
    \includegraphics[width=0.185\linewidth]{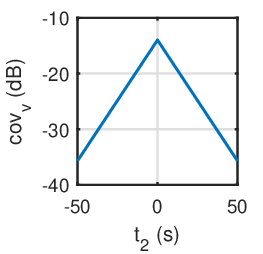}
    \label{fig:WindCovSlice_dB_WSSexp_Tv10.0_mu0.50_sig0.20}}
    \hfill
  \subfloat[]{
    \includegraphics[width=0.185\linewidth]{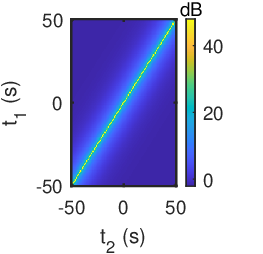}
    \label{fig:Rh2D_dB_tau4.00_WSSexp_Tv10.0_mu0.50_sig0.20_Rx_[0.71_0.71_1.00]}}
    \hfill
  \subfloat[]{
    \includegraphics[width=0.185\linewidth]{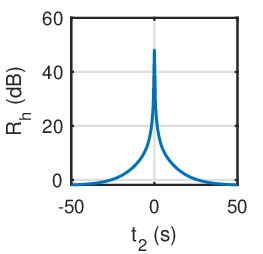}
    \label{fig:RhSlice_dB_tau4.00_WSSexp_Tv10.0_mu0.50_sig0.20_Rx_[0.71_0.71_1.00]}}
  \\
  \subfloat[]{
    \includegraphics[width=0.185\linewidth]{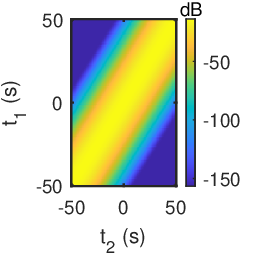}
    \label{fig:WindCov2D_dB_WSSgauss_Tv10.0_mu0.50_sig0.20}}
    \hfill
  \subfloat[]{
    \includegraphics[width=0.185\linewidth]{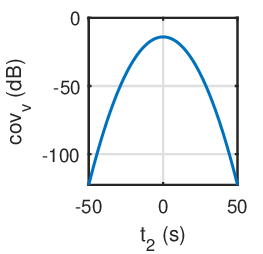}
    \label{fig:WindCovSlice_dB_WSSgauss_Tv10.0_mu0.50_sig0.20}}
    \hfill
  \subfloat[]{
    \includegraphics[width=0.185\linewidth]{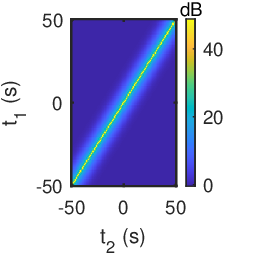}
    \label{fig:Rh2D_dB_tau4.00_WSSgauss_Tv10.0_mu0.50_sig0.20_Rx_[0.71_0.71_1.00]}}
    \hfill
  \subfloat[]{
    \includegraphics[width=0.185\linewidth]{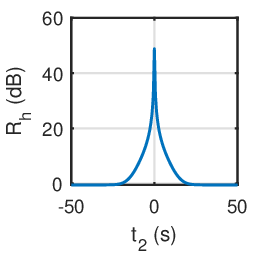}
    \label{fig:RhSlice_dB_tau4.00_WSSgauss_Tv10.0_mu0.50_sig0.20_Rx_[0.71_0.71_1.00]}}
    \\
    
    \caption{\(R_h(\tau_1 =\tau_2 =4;t_1,t_2)\) plots for different wind setups: 
    \protect\subref{fig:WindCov2D_dB_WSSexp_Tv10.0_mu0.50_sig0.20} WSS Exponential wind covariance and
    \protect\subref{fig:WindCovSlice_dB_WSSexp_Tv10.0_mu0.50_sig0.20} its slice at \(t_1 = 0\),   
    \protect\subref{fig:Rh2D_dB_tau4.00_WSSexp_Tv10.0_mu0.50_sig0.20_Rx_[0.71_0.71_1.00]} corresponding channel autocorrelation and
    \protect\subref{fig:RhSlice_dB_tau4.00_WSSexp_Tv10.0_mu0.50_sig0.20_Rx_[0.71_0.71_1.00]} its slice at \(t_1 = 0\);
    \protect\subref{fig:WindCov2D_dB_WSSgauss_Tv10.0_mu0.50_sig0.20} WSS Gaussian wind covariance and
    \protect\subref{fig:WindCovSlice_dB_WSSgauss_Tv10.0_mu0.50_sig0.20} its slice at \(t_1 = 0\), 
    \protect\subref{fig:Rh2D_dB_tau4.00_WSSgauss_Tv10.0_mu0.50_sig0.20_Rx_[0.71_0.71_1.00]} corresponding channel autocorrelation and
    \protect\subref{fig:RhSlice_dB_tau4.00_WSSgauss_Tv10.0_mu0.50_sig0.20_Rx_[0.71_0.71_1.00]} its slice at \(t_1 = 0\).}
    
  \label{Wind_channel_autocov_pairs}
\end{figure*}

\begin{figure}[!t]
  \centering

  \subfloat[]{
    \includegraphics[width=0.45\linewidth]{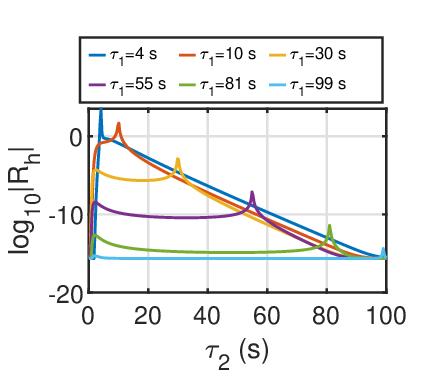} 
    \label{fig:Rh_slice_vs_tau2_WSSGauss_Tv10.0_mu0.50_sig0.20_Rx_[0.71_0.71_1.00]_t1_120.00_t2_120.00_taus_4_10_30_55_81_99}}
  \subfloat[]{
    \includegraphics[width=0.45\linewidth]{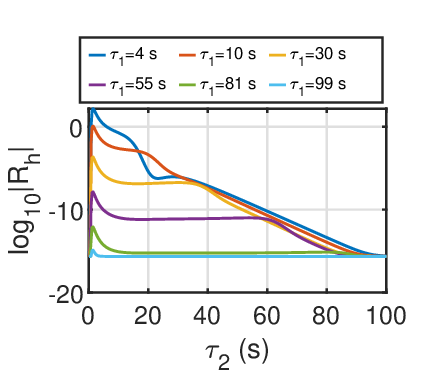}
    \label{fig:Rh_slice_vs_tau2_WSSGauss_Tv10.0_mu0.50_sig0.20_Rx_[0.71_0.71_1.00]_t1_120.00_t2_130.00_taus_4_10_30_55_81_99}}
  \\
  \subfloat[]{
    \includegraphics[width=0.45\linewidth]{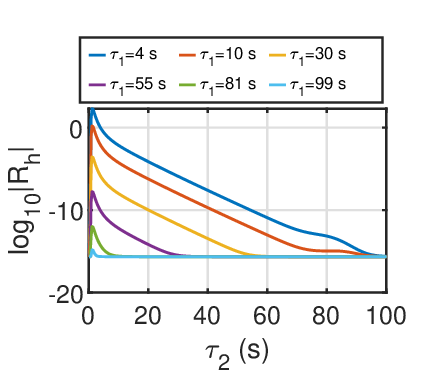}
    \label{fig:Rh_slice_vs_tau2_WSSGauss_Tv10.0_mu0.50_sig0.20_Rx_[0.71_0.71_1.00]_t1_120.00_t2_200.00_taus_4_10_30_55_81_99}}

  \caption{Logarithm of \(R_{h}\) vs \(\tau_2\) for different \(\tau_1\) values where \(t_1 = 120\)
  \protect\subref{fig:Rh_slice_vs_tau2_WSSGauss_Tv10.0_mu0.50_sig0.20_Rx_[0.71_0.71_1.00]_t1_120.00_t2_120.00_taus_4_10_30_55_81_99} \(t_2 = 120\), 
  \protect\subref{fig:Rh_slice_vs_tau2_WSSGauss_Tv10.0_mu0.50_sig0.20_Rx_[0.71_0.71_1.00]_t1_120.00_t2_130.00_taus_4_10_30_55_81_99} \(t_2 = 130\), and 
  \protect\subref{fig:Rh_slice_vs_tau2_WSSGauss_Tv10.0_mu0.50_sig0.20_Rx_[0.71_0.71_1.00]_t1_120.00_t2_200.00_taus_4_10_30_55_81_99} \(t_2 = 200\).}
  \label{R_h_vs_tau2_different_time_seperations}
\end{figure}

The only time dependent components in the autocorrelation of \(h(\tau,t) \) are \(\sigma^{2}_{X}(\tau,t)\) and \(L(\tau_{1},\tau_{2};\, t_{1},t_{2})\). Using (\ref{eq:sigmaX_definition}) and (\ref{eq:L_definition}), it can be shown that when the wind is WSS, the channel also becomes WSS, so it can be represented as \(R_h(\tau_1,\tau_2;\Delta t)\). For different definitions of wind covariance, the resulting channel autocorrelation is plotted for \(\tau_1 = \tau_2 = 4 \, s\). In this analysis, the point source is located at \(\mathbf{z_0} = [0 \, 0 \, 1]\) and the detection point is located at \(\mathbf{x} = [ \frac{\sqrt{2}}{2} \, \frac{\sqrt{2}}{2}  \,1 ]\) so that the mean wind flows from the source to the detection point. In addition, the diffusion coefficient of (Z)-3-hexenyl acetate \(D = 6.7698 \times 10^{-6} \, m^2/s\) is used as a representative value. This specific molecule is released into the air from plants under herbivore attack and used as an information carrier \cite{The_Carboxylesterase_AtCXE12_Converts_Volatile_Z3Hexenyl_Acetate_to_Z3Hexenol_in_Arabidopsis_Leaves}. Two different wind covariances are used, which are WSS Exponential wind covariance \(C_v(t_1, t_2) = (0.2)^2 \exp( - \frac{|t_1 - t_2|}{10})\) and WSS Gaussian wind covariance \( C_v(t_1, t_2) = (0.2)^2 \exp( - \left( \frac{t_1 - t_2}{10} \right)^2) \). The resulting plots are shown in  \autoref{Wind_channel_autocov_pairs}.The channel acts as a decorrelator after a small time difference for a constant delay.
Using the Gaussian WSS covariance for the wind with a mean of \(0.5 \, m/s\) and a standard deviation of \(0.2 \, m/s\) shown in \autoref{Wind_channel_autocov_pairs}-\subref{fig:WindCov2D_dB_WSSgauss_Tv10.0_mu0.50_sig0.20}, the logarithm of \(|R_{h}(\tau_{1},\tau_{2};\, t_{1},t_{2})|\) is plotted against \(\tau_2 \) for different values of \(\tau_1\) and time separation. The resulting plot is given in \autoref{R_h_vs_tau2_different_time_seperations}. As the time separation increases, the autocorrelation between different delays drops drastically. Moreover, in \autoref{R_h_vs_tau2_different_time_seperations}-\subref{fig:Rh_slice_vs_tau2_WSSGauss_Tv10.0_mu0.50_sig0.20_Rx_[0.71_0.71_1.00]_t1_120.00_t2_120.00_taus_4_10_30_55_81_99}, for a time separation small enough, the autocorrelation again decreases for a delay separation around \(5\) seconds. \autoref{Wind_channel_autocov_pairs} and \autoref{R_h_vs_tau2_different_time_seperations} use colored WSS Gaussian wind only to illustrate the effect of wind covariance on the channel autocorrelation. All subsequent analyses assume independent WSS white Gaussian wind components.

\subsection{Comparison of diffusion-advection and Conventional Time-varying Channels}

The similarity between the classical wireless channel and the diffusion-advection channel can be first observed from (\ref{eq:c_solution_point_source_phiD}). For both channels, a time-varying impulse response is defined. In general, the wireless channels are studied under the wide-sense stationary uncorrelated scattering (WSSUS) assumption. However, the assumption of uncorrelated scattering is not suitable for diffusion-advection channels. This is because particles in the air collide with each other and diffuse together. The scatterers in this problem are the information carriers at the same time. In this paper, the comparison between wireless and diffusion-advection channels is conducted using the Power Delay Profile (PDP) of these channels.

\begin{figure}[!t]
  \centering

  \subfloat[]{
    \includegraphics[width=0.40\linewidth]{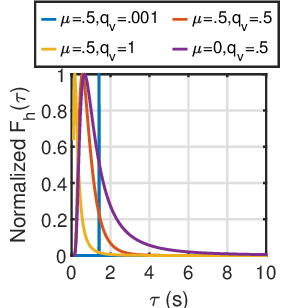} 
    \label{fig:PDP_tau_Rx_[0.71_0.71_1.00]_mu_[0.500_0.500_0.500_0.000]_sig_[0.001_0.500_1.000_0.500]}}
  \subfloat[]{
    \includegraphics[width=0.40\linewidth]{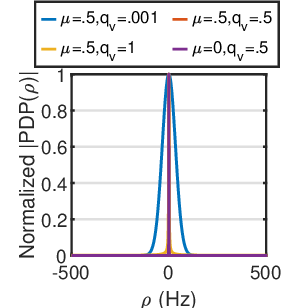}
    \label{fig:FFT_PDP_Rx_[0.71_0.71_1.00]_mu_[0.500_0.500_0.500_0.000]_sig_[0.001_0.500_1.000_0.500]}}
  \\
  \subfloat[]{
    \includegraphics[width=0.40\linewidth]{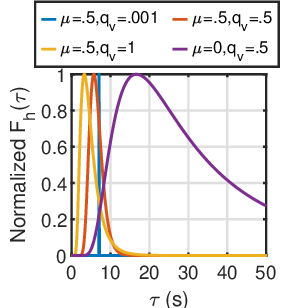}
    
    \label{fig:PDP_tau_Rx_[3.54_3.54_1.00]_mu_[0.500_0.500_0.500_0.000]_sig_[0.001_0.500_1.000_0.500]}}
  \subfloat[]{
    \includegraphics[width=0.40\linewidth]{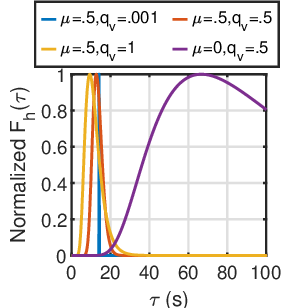}
    \label{fig:PDP_tau_Rx_[7.07_7.07_1.00]_mu_[0.500_0.500_0.500_0.000]_sig_[0.001_0.500_1.000_0.500]}}

  \caption{Power Delay Profile \(F_h(\tau)\) of the channel for different wind assumptions at distances of
  \protect\subref{fig:PDP_tau_Rx_[0.71_0.71_1.00]_mu_[0.500_0.500_0.500_0.000]_sig_[0.001_0.500_1.000_0.500]} \(1 \, meter\), 
  \protect\subref{fig:PDP_tau_Rx_[3.54_3.54_1.00]_mu_[0.500_0.500_0.500_0.000]_sig_[0.001_0.500_1.000_0.500]} \(5 \, meters\), 
  \protect\subref{fig:PDP_tau_Rx_[7.07_7.07_1.00]_mu_[0.500_0.500_0.500_0.000]_sig_[0.001_0.500_1.000_0.500]} \(10 \, meters\) and 
  \protect\subref{fig:FFT_PDP_Rx_[0.71_0.71_1.00]_mu_[0.500_0.500_0.500_0.000]_sig_[0.001_0.500_1.000_0.500]} the Fourier transforms of \(F_h(\tau)\) for  \protect\subref{fig:PDP_tau_Rx_[0.71_0.71_1.00]_mu_[0.500_0.500_0.500_0.000]_sig_[0.001_0.500_1.000_0.500]}.  }
  \label{Power_delay_profiles}
\end{figure}

The PDP \(F_h(\tau)\) of the diffusion-advection channel is obtained by inserting \(\Delta t = 0\) and \(\tau_1 = \tau_2 = \tau\) into the WSS channel autocorrelation function \(R_{h}(\tau_{1},\tau_{2};\, \Delta t)\). Four representative wind conditions are examined: \textit{(i)} dominant wind mean, \textit{(ii)} comparable wind mean and $q_v$, \textit{(iii)} dominant wind $q_v$, and \textit{(iv)} zero wind mean. For each case, PDP is computed at three source–detector separations (\(1 \, m\), \(5 \, m\), \(10 \, m\)) assuming that the mean wind is directed from the source to the detector. In addition, the Fourier transform of the PDP is evaluated only for the \(1 \, \text{m}\) separation as a representative case to illustrate how channel dispersion manifests in the frequency domain, since similar trends are observed for larger distances. These results are presented in \autoref{Power_delay_profiles}. When the wind mean dominates, a very narrow peak is observed in the PDP in the time domain. Increasing the distance shifts when this peak occurs. As $q_v$ increases compared to the wind mean, the time interval in which the PDP remains appreciable increases, resulting in a narrower bandwidth in the frequency domain. This means that the channel loses its ability to decorrelate the released particles. When the wind mean is not directed from source to detector, or the wind mean is zero, establishing a communication system where bit-by-bit transmission is employed is not reliable. There is no guarantee that the particles will ever reach the receiver, and even if some particles are detected at the receiver, currently, there is no algorithm that can enable it to decide which particles were released first. This is because the particles in the air remain correlated for a very long time. With the mean wind directed from the source to the detector, a bit-wise communication link can be established. In this topology, the transmitter sends symbols using pulses of a pre-determined duration, and the receiver reads these pulses after they pass through the channel. This is parallel to what happens in a wireless channel. In wireless communications, the RMS delay spread is commonly used to quantify temporal dispersion by measuring the statistical spread of the PDP around its mean delay. We adopt the same second-moment measure for the diffusion-advection channel and refer to it as the channel dispersion time \(T_d\), emphasizing its connection to the physical spreading of particles. While the first-arrival time primarily represents propagation delay determined by the mean wind and transmitter-receiver separation, \(T_d\) characterizes the temporal dispersion that governs pulse overlap, inter-symbol interference, and the selection of an appropriate symbol duration. Its formal definition is presented in the next section.

\subsection{Channel Dispersion Time}
\begin{figure}[!h]
  \centering
  \includegraphics[width=0.6\linewidth]{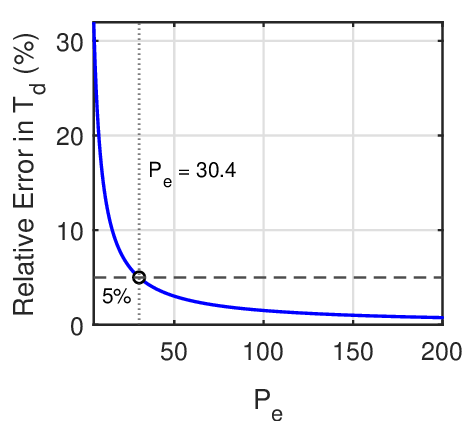}
  \caption{Relative error of the high-$P_e$ channel dispersion time approximation in \eqref{eq:T_d_definition} with respect
  to the reduced Bessel expression in
  \eqref{eq:rms_bessel_expression}.}
  \label{fig:RMS_delay_spread_relative_error}
\end{figure}

The channel dispersion time \(T_d\), defined as the RMS delay spread, quantifies the statistical spread of the PDP around its mean delay. The PDP of the channel is written as
\begin{equation}
\label{eq:power_del_prof_expression}
\resizebox{0.95\linewidth}{!}{$
\begin{aligned}
F_{h}(\tau)
&= \beta^{2}(\tau)\,
   \frac{D}{D_{eff}}\,
   \exp\!\left(
   -\frac{
      (x_{1} - z_{01} - \mu \tau)^{2}
      + (x_{2} - z_{02} - \mu \tau)^{2}
   }{
      2\,\tau\,D_{eff}
   }
   \right),
\end{aligned}
$}
\end{equation}
\noindent where $D_{eff} = D + \frac{\sigma_X^2}{\tau}$ ($ m^2/s$) can be considered as an effective diffusion coefficient. For white Gaussian wind with covariance $C_v(t,t')=q_v^2\delta(t-t')$, we have $D_{\mathrm{eff}}=D+q_v^2$, where $q_v^2$ has units of $\mathrm{m^2/s}$. Thus, increasing $q_v^2$ increases $D_{eff}$. Determination of \(T_d\) is related to the P\'eclet number (\(P_e\)), which characterizes whether advection or diffusion dominates the flow \cite{Modeling_Duct_Flow_for_Molecular_Communication}. It can be written as
\begin{equation}
\label{eq:Peclet_number}
\begin{split}
P_e = \frac{\text{advection rate}}{\text{diffusion rate}} = \frac{L \sqrt{2}\mu}{D_{eff}},
\end{split}
\end{equation}
\noindent where \(L = \sqrt{(x_{1} - z_{01})^{2}+ (x_{2} - z_{02})^{2}}\) is the distance between the transmitter and receiver, assuming that both are in the same height level. For sufficiently large $P_e$, advection dominates and, as derived from the PDP moments in the Appendix, the channel dispersion time has the asymptotic form
\begin{equation}
\label{eq:T_d_definition}
\begin{aligned}
T_d
&\sim
\sqrt{
\frac{
LD_{eff}
}{
2\sqrt{2}\,\mu^3
}
},
\qquad
P_e\rightarrow\infty.
\end{aligned}
\end{equation}
The accuracy of \eqref{eq:T_d_definition} is evaluated relative to the reduced Bessel expression in \eqref{eq:rms_bessel_expression}. As shown in \autoref{fig:RMS_delay_spread_relative_error}, the relative error in $T_d$ decreases rapidly as $P_e$ increases, confirming the validity of the approximation for sufficiently large $P_e$.

\(T_d\) is a time-scale that characterizes how long the channel impulse response remains significantly spread in delay due to the combined effects of diffusion and stochastic advection. It depends on the separation between the transmitter and the receiver because increasing the distance increases the particle transit time, during which diffusion and random wind fluctuations accumulate. Using (\ref{eq:T_d_definition}) and a symbol duration of \(T_{\mathrm{sym}}\), the diffusion-advection channel can be classified as
\[
\begin{aligned}
T_{\mathrm{sym}} \geq T_{d} &:\ \text{Non-dispersive channel}, \\
T_{\mathrm{sym}} < T_{d} &:\ \text{Dispersive channel}.
\end{aligned}
\]

\begin{figure*}[t]
    \centering
    \includegraphics[width=\textwidth]{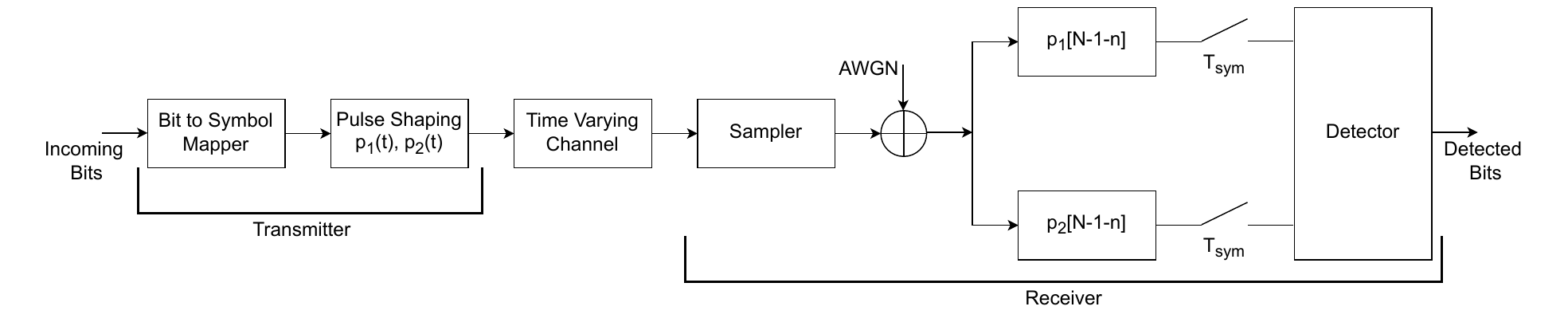}
    \caption{Block diagram of the complete communication link, including transmitter processing, pulse-shaped emission through the time-varying diffusion-advection channel, sampling with AWGN, and the receiver detection stage.}
    \label{Primitive_system_block_diagram_V2}
\end{figure*}
For symbol durations exceeding the channel dispersion time, the dominant portion of the channel impulse response is confined within a single symbol interval, resulting in reduced inter-symbol interference. In contrast, when the symbol duration is shorter than the dispersion time, significant portions of the impulse response extend beyond one symbol period, leading to persistent ISI. The channel dispersion time is introduced as a design-relevant time scale that quantifies the effective memory of the diffusion–advection channel. It provides a guideline for selecting symbol durations such that the majority of the channel-induced temporal spreading is contained within a single symbol interval. It follows from (\ref{eq:T_d_definition}) that increasing $q_v$ increases the channel dispersion time, thereby extending the effective channel memory and necessitating longer symbol durations to limit ISI. Conversely, increasing the mean wind speed \(\mu\) reduces the dispersion time by accelerating coherent transport, which confines the impulse response in time and enables higher symbol rates. This distinction reveals that reliable airborne particle communication requires directed advection, where transport is dominated by a consistent mean flow component. Under such directed flow conditions, particles experience reduced temporal spreading during propagation. This regime therefore enables a predictable arrival structure and bounded channel memory, which are essential for reliable pulse-based communication. Conversely, advection characterized by high $q_v$ and a weak mean flow lacks directional dominance, resulting in increased dispersion.

\begin{figure}[!h]
  \centering
  \includegraphics[width=0.6\linewidth]{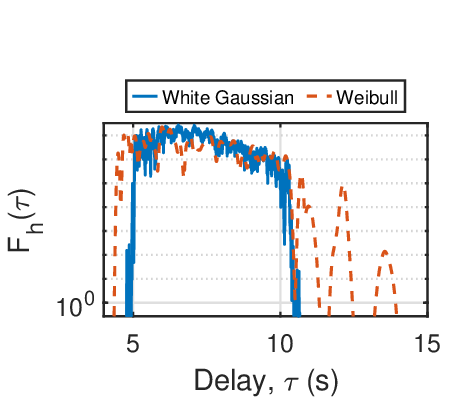}
  \caption{Time-averaged PDPs obtained under the white
  Gaussian and correlated signed-Weibull wind models.}
  \label{fig:wind_model_sensitivity_PDP}
\end{figure}

To assess the sensitivity of the channel results to the white-Gaussian wind assumption, we consider a correlated non-Gaussian wind model. Following Weibull-based joint models of wind speed and direction \cite{Joint_Modeling_of_Wind_Speed_and_Wind_Direction_Through_a_Conditional_Approach}, we represent each wind component as \(V=BS\), where \(S\sim\operatorname{Weibull}(k,\lambda)\), \(\Pr(B=+1)=p\), and \(\Pr(B=-1)=1-p\). An empirical analysis of ERA-40 wind data reports fitted Weibull shape parameters in the range \(k\in[1.4,2.6]\) \cite{Comprehensive_Empirical_Analysis_of_ERA40_Surface_Wind_Speed_Distribution_over_Europe}. We therefore select \(k=2.5\) and determine \(\lambda=0.11585~\mathrm{m/s}\) and \(p=0.98644\) to obtain \(\mathbb{E}[V]=0.1~\mathrm{m/s}\) and \(\operatorname{Std}[V]=0.05~\mathrm{m/s}\). Temporal dependence is specified through the normalized exponential autocorrelation function \(\rho_V(\Delta)=\exp(-|\Delta|/T_V)\), following its established use in non-Gaussian wind modeling \cite{Construction_of_SDE_based_wind_speed_models_with_exponentially_decaying_autocorrelation}. We adopt an integral time scale of \(2~\mathrm{s}\), representative of low-speed indoor airflow \cite{The_measurement_and_simulation_of_indoor_air_flow}. Since the integral time scale of the selected autocorrelation function equals \(T_V\), we set \(T_V=2~\mathrm{s}\). The correlated signed-Weibull samples are generated using the ARTA method \cite{Autoregressive_to_anything_Time_series_input_processes_for_simulation,Numerical_Methods_for_Fitting_and_Simulating_Autoregressive_to_Anything_Processes}. As shown in \autoref{fig:wind_model_sensitivity_PDP}, the correlated signed-Weibull model produces a broader and more irregular PDP. The corresponding channel dispersion time increases from \(T_d=0.992~\mathrm{s}\) for the white Gaussian model to \(T_d=1.367~\mathrm{s}\), representing an increase of \(37.8\%\). Despite its sensitivity to wind statistics, the Gaussian model provides a useful initial estimate of \(T_d\) for assessing temporal dispersion and selecting appropriate symbol durations.

\section{Communication Under Directed Wind}
\subsection{Particle Transmitter}

The release of particles into the air is a simple procedure. Basically, any type of spray can be considered a possible transmission device. In digital communications, after the information bits are mapped to the designated symbols, application-engineered pulses are used in transmission. The same signaling structure is adopted in this paper. The natural way to model particle transmission in time is by using rectangular pulses. The symbol period and the number of orthogonal pulses being used constitute the parameters of the particle pulse shaping design. System shown in \autoref{Primitive_system_block_diagram_V2} uses two orthogonal rectangle pulses with a symbol period of \(T_{sym} = 2\) seconds. The first pulse \(p_1(t)\) is \(1\) for \( t \in [0,1) \) and 0 for \( t \in [1,2) \). The second pulse \(p_2(t)\) is the complement of the first one. Therefore, two orthogonal unit energy pulses are utilized in this configuration.  The simulations were performed in MATLAB R2023b.

\subsection{Particle Receiver}

Gas detectors operate by converting chemical cues into measurable electronic signals, enabling the detection and quantification of various concentrations of gas particles in the air \cite{Comprehensive_Review_on_Gas_Sensors_Unveiling_Recent_Developments_and_Addressing_Challenges}, \cite{Gas_Sensors_A_Review}. A short overview of current gas detectors and their working principles is given in \autoref{Gas_sensor_Types}. In the context of molecular communications, most studies use MOS gas sensors, since they are simple and easy to experiment with. However, these sensors have a response time of seconds to minutes, which is too slow \cite{HighBandwidth_eNose_for_Rapid_Tracking_of_Turbulent_Plumes}. Although there are methods to use these sensors in a faster manner, these methods are specific to the detection of bit 1 or 0 \cite{Resolving_Fast_Gas_Transients_with_Metal_Oxide_Sensors}. The fastest gas sensors are optical sensors with a response time of milliseconds \cite{Response_Time_for_Optical_Emission_and_Mass_Spectrometric_Signals_During_Etching_of_Heterostructures}. These sensors can also operate at room temperature and are tunable to a specific gas particle \cite{Exhaled_Breath_Analysis_Through_the_Lens_of_Molecular_Communication_A_Survey}. If the subtractive demodulation algorithm is used, the receiver becomes linear \cite{Improvement_of_the_Detection_Sensitivity_for_Tunable_Diode_Laser_Absorption_Spectroscopy_A_Review}. It is assumed that the receiver takes many samples from the air in one symbol duration, and matched filtering is applied in the digital domain. The sampling rate of the receiver is taken as \(0.1\) \(kHz\) assuming a linear optical gas sensor is utilized. Moreover, for simplicity of the analysis, the receiver is assumed to directly measure the air concentration without any scaling. Regarding receiver noise, we assume a thermal-noise-limited regime \cite{Signal_to_noise_ratio_analysis_in_laser_absorption_spectrometers_using_optical_multipass_cells} and model the sampled noise as additive white Gaussian noise.

\begin{table*}[t]
\centering
\caption{Summary of Gas Sensor Types and Operating Principles}
\label{Gas_sensor_Types}
\footnotesize 
\renewcommand{\arraystretch}{1.3} 

\begin{tabular}{p{0.13\textwidth} p{0.32\textwidth} p{0.13\textwidth} p{0.32\textwidth}}
\toprule
\textbf{Sensor Type} & \textbf{Working Principle} & \textbf{Sensor Type} & \textbf{Working Principle} \\
\midrule

\textbf{Catalytic Gas Sensor} 
& Catalytic combustion increases the temperature of a platinum element, changing its resistance \cite{A_High_Heating_Efficiency_TwoBeam_Microhotplate_for_Catalytic_Gas_Sensors}.
& \textbf{Polymer-Based Gas Sensor} 
& Gas–polymer interaction alters electrical/chemical properties of the polymer layer \cite{A_Review_of_Composite_Conducting_PolymerBased_Sensors_for_Detection_of_Industrial_Waste_Gases}. \\

\textbf{Optical Gas Sensor} 
& Measures gas-induced variations in light absorption, reflection, or scattering \cite{Optical_Gas_Sensing_A_Review}.
& \textbf{Carbon Nanotube (CNT) Gas Sensor} 
& Gas exposure causes charge transfer with nanotubes, modifying their electrical conductivity \cite{An_LC_Passive_Wireless_Gas_Sensor_Based_on_PANICNT_Composite}. \\

\textbf{Electrochemical Gas Sensor} 
& Target gas undergoes redox reaction; resulting current is proportional to concentration \cite{A_Wearable_Electrochemical_Gas_Sensor_for_Ammonia_Detection}.
& \textbf{MOS Gas Sensor} 
& Gas adsorption changes the conductivity of metal oxide semiconductor material \cite{Highly_Sensitive_and_Selective_Gas_Sensors_Using_PType_Oxide_Semiconductors_Overview}. \\

\textbf{Thermal Conductivity Gas Sensor} 
& Measures heat loss from a heated element to surrounding gas; conductivity varies with gas type \cite{Design_of_Thermal_Conductivity_Gas_Sensor_with_Silicon_Cap}.
& \textbf{Schottky Diode Gas Sensor} 
& Gas interaction forms a dipole layer at the metal–semiconductor interface, altering Schottky barrier height \cite{Simplified_Gas_Sensor_Model_Based_on_AlGaNGaN_Heterostructure_Schottky_Diode}. \\

\textbf{Infrared Gas Sensor} 
& Detects gas by IR absorption at characteristic molecular vibration wavelengths \cite{A_Design_of_an_UltraCompact_Infrared_Gas_Sensor_for_Respiratory_Quotient_qCO2_Detection}.
& \textbf{MEMS Gas Sensor} 
& Gas adsorption influences MEMS mechanical/electrical properties (resonance, deflection, conductivity) \cite{Mass_Analysis_of_CH4SO2_Gas_Mixture_by_LowPressure_MEMS_Gas_Sensor}. \\

\textbf{Acoustic Wave Gas Sensor} 
& Gas particles interact with the sensing layer and change its properties, altering the propagation of an acoustic wave traveling along the sensor structure \cite{Miniaturized_Thermal_Acoustic_Gas_Sensor_Based_on_a_CMOS_Microhotplate_and_MEMS_Microphone}.
& \textbf{Magnetic Gas Sensor} 
& Gas exposure modifies magnetic properties of the sensing material \cite{Magnetic_Gas_Sensing_Working_Principles_and_Recent_Developments}. \\

\bottomrule
\end{tabular}
\end{table*}

\subsection{Simulation Results for Directed Wind}

For simulations, both pulses \(p_1(t)\) and \(p_2(t)\) are created with a sampling rate of \(0.1\) \(kHz\) in the transmitter to match them exactly in the receiver. Before being applied to the channel, the signals are expanded onto a \(1~\mathrm{kHz}\) simulation grid using a zero-order hold, and the channel is simulated on the same grid. Unless otherwise specified in the corresponding figure caption, \(30~\mathrm{s}\) of channel memory is considered to reduce computational cost. Numerical integration of the PDP showed that the maximum channel-energy fraction beyond \(30~\mathrm{s}\) is \(8.42\times10^{-24}\) across all considered simulation settings. Thus, the selected memory retains all numerically significant channel energy. The transmitter is located at \(\mathbf{z_0} = [0 \, 0 \, 1]\) and the receiver is located at \(\mathbf{x} = [ \frac{\sqrt{2}}{2} \, \frac{\sqrt{2}}{2}  \,1 ]\) so that a positive mean wind indicates a mean flow from the transmitter to the receiver.

In this system, the receiver samples the concentration of the particle to which it is calibrated all the time. To determine when signaling begins and overcome any symbol timing offset, pilot signals are utilized. In the receiver, a two-step approach is used to find the correct symbol timing. An initial index is determined by scanning the total energy of the output of the matched filters in memory. In the first step, the first index that exceeds \(0.05\) times the total energy of the output of the matched filters is chosen. In the second step, all possible symbol detection times are tested around the first index, and the inner product between the observed matched filter outputs and the expected symbol shape is calculated. The index maximizing this inner product is chosen as the initial sampling time for detection. Then, consecutive samplings are performed with period \(T_{sym}\). To mitigate the channel scaling effect, an MMSE equalizer is used in the signal space. This choice is motivated by the strong directed advection regime, where the power delay profile is sharply concentrated, and the channel behaves primarily as a scaling with limited dispersion. In contrast, diffusion-dominated channels exhibit a more spread-out power delay profile and increased dispersion, making them less suitable for linear techniques such as MMSE. Therefore, while MMSE equalization is effective in the advection-dominated case, diffusion-dominated channels may require nonlinear detection techniques. Exploring such techniques constitutes an important direction for future work. The coefficients of the MMSE equalizer are determined using the pilot signals. Moreover, the additive white Gaussian noise is defined at the output of the sampled channel. In the proposed system, the received concentration is converted into an electrical signal and then processed digitally; thermal noise in the electronic circuitry can consequently produce negative samples. The equalizer output is an electrical-domain decision statistic rather than a concentration, so negative values after equalization do not imply negative particle concentrations. Detection is performed using the maximum likelihood criterion that boils down to the minimum Euclidean distance. The simulation begins with the first symbol transmission, and additional empty symbols are appended after the actual data sequence to allow observation of the channel response after the transmission has ended.

\begin{figure}[!t]
  \centering

  \subfloat[]{
    \includegraphics[width=0.28\linewidth]{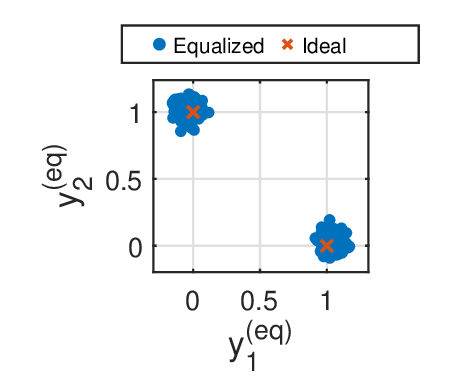} 
    \label{fig:Comm_Equalized_Constellation_SNR_10dB_mu_0.500_sig_0.001_mod_1_Tsym_2.000}}
  \hspace{0.015\linewidth}
  \subfloat[]{
    \includegraphics[width=0.28\linewidth]{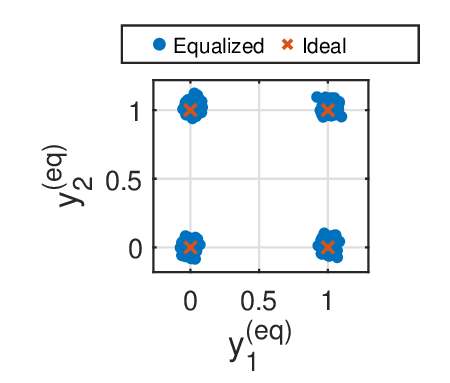}
    \label{fig:Comm_Equalized_Constellation_SNR_10dB_mu_0.500_sig_0.001_mod_2_Tsym_2.000}}
  \hspace{0.015\linewidth}
  \subfloat[]{
    \includegraphics[width=0.28\linewidth]{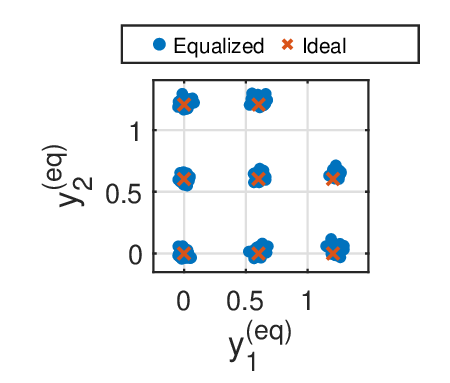}
    \label{fig:Comm_Equalized_Constellation_SNR_10dB_mu_0.500_sig_0.001_mod_3_Tsym_2.000}}
  \\
  \subfloat[]{
    \includegraphics[width=0.28\linewidth]{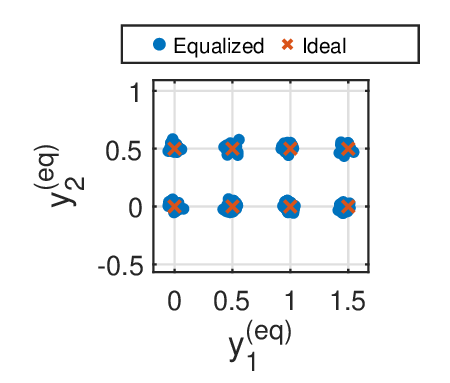}
    \label{fig:Comm_Equalized_Constellation_SNR_10dB_mu_0.500_sig_0.001_mod_4_Tsym_2.000}}
\hspace{0.015\linewidth}
  \subfloat[]{
    \includegraphics[width=0.28\linewidth]{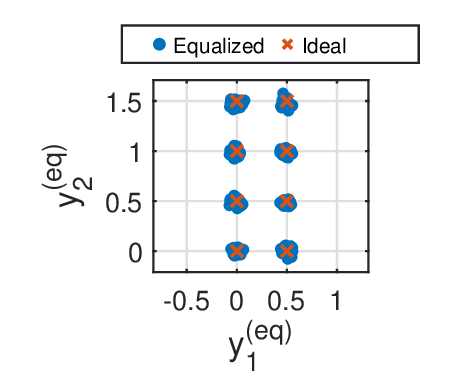}
    \label{fig:Comm_Equalized_Constellation_SNR_10dB_mu_0.500_sig_0.001_mod_5_Tsym_2.000}}
  \hspace{0.015\linewidth}
  \subfloat[]{
    \includegraphics[width=0.28\linewidth]{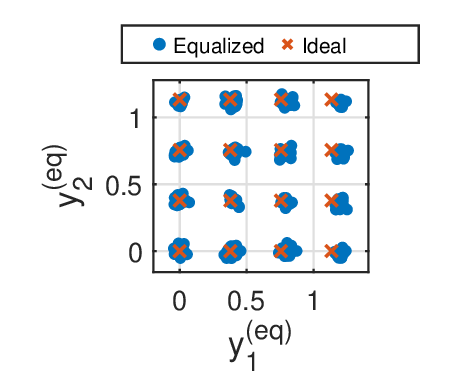}
    \label{fig:Comm_Equalized_Constellation_SNR_10dB_mu_0.500_sig_0.001_mod_6_Tsym_2.000}}
    
    \caption{Constellation diagrams for 
    \protect\subref{fig:Comm_Equalized_Constellation_SNR_10dB_mu_0.500_sig_0.001_mod_1_Tsym_2.000} 2-symbol modulation,
    \protect\subref{fig:Comm_Equalized_Constellation_SNR_10dB_mu_0.500_sig_0.001_mod_2_Tsym_2.000} 4-symbol modulation,
    \protect\subref{fig:Comm_Equalized_Constellation_SNR_10dB_mu_0.500_sig_0.001_mod_3_Tsym_2.000} 8-symbol symmetric modulation,
    \protect\subref{fig:Comm_Equalized_Constellation_SNR_10dB_mu_0.500_sig_0.001_mod_4_Tsym_2.000} 8-symbol wide modulation, 
    \protect\subref{fig:Comm_Equalized_Constellation_SNR_10dB_mu_0.500_sig_0.001_mod_5_Tsym_2.000} 8-symbol tall modulation and 
    \protect\subref{fig:Comm_Equalized_Constellation_SNR_10dB_mu_0.500_sig_0.001_mod_6_Tsym_2.000} 16-symbol modulation.}
    
  \label{Constellation_plots}
\end{figure}

Since the signal is represented by the concentrations of particles, only the part where pulses have nonnegative coefficients can be used in the signal space. Using the two defined orthogonal pulses, six different modulation schemes are inspected. Taking \(\mu = 0.5 \,m/s\), \(q_v = 3.16 \times 10^{-5} \,m/\sqrt{s}\), \(SNR = 10 \,dB\) and \(T_{sym} = 2 \, s\), 300 symbols were transmitted using 10 pilot symbols for each modulation type, and the resulting constellations are given in \autoref{Constellation_plots}. The simulation results indicate that the receiver is capable of decoding the transmitted signals successfully. This is a promising result, since most studies on molecular communication are limited to using only two different symbols per molecule type, due to the restrictive characteristics of diffusive channels. With controlled advection, one can utilize multi-symbol constellations with a single particle type. This demonstrates that the communication space is not inherently binary, but can be structured through pulse shaping and receiver processing. This shifts the design space from molecule-limited modulation to waveform-limited modulation, where symbol design is governed by channel dispersion and receiver resolution. This can change how we engineer particle-based communications for real-life applications. For example, higher data rates can be achieved without increasing the chemical complexity of the system.

\begin{figure}[!h]
  \centering
  \includegraphics[width=0.6\linewidth]{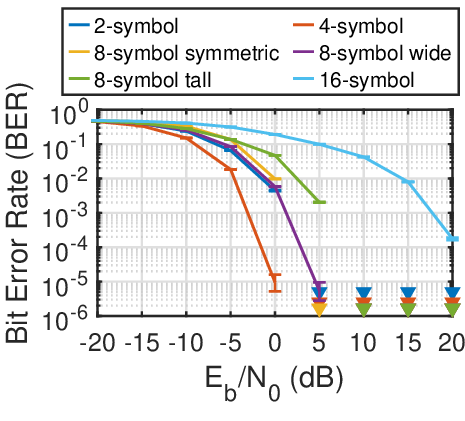}
  \caption{BER performance of the six modulation schemes as a function of $E_b/N_0$. Results are averaged over 736 Monte Carlo trials. For the employed channel memory of \(T_{\mathrm{mem}}=3\) s, the omitted channel-energy fraction is below double-precision numerical resolution. Downward triangles denote 95\% upper confidence bounds for zero-error cases.}
  \label{BER_EbN0_all_modtypes_COMBINED}
\end{figure}

To compare the performance of these modulations, the BER-\(E_b/N_0\) plots are obtained. In this analysis, the AWGN was defined with respect to the transmitter's average symbol and pulse energies. Moreover, for the signal-noise amplitude match, the channel was normalized to unit energy only for this analysis. \(\mu = 0.5 \,m/s\), \(q_v = 3.16 \times 10^{-5} \,m/\sqrt{s}\), and \(T_{sym} = 2 \, s\) were set. 1000 data symbols and 10 pilot symbols were transmitted for each modulation type. The results are shown in \autoref{BER_EbN0_all_modtypes_COMBINED}. Interestingly, the results show that the optimal modulation is not the lowest-order scheme, but the 4-symbol constellation. This indicates that particle-based channels with controlled advection can benefit from moderate constellation expansion, which improves symbol distinguishability without excessively amplifying noise sensitivity. However, in general, it is observed that as the cardinality of the modulation increases, the performance decreases. It is emphasized that the presented BER results are based on numerical simulations and serve to illustrate the potential benefits and limitations of higher-order modulation in diffusion–advection channels. Confirming these trends under different channel realizations and through experimental studies constitutes an important direction for future work.

\begin{figure}[!h]
  \centering
  \includegraphics[width=0.6\linewidth]{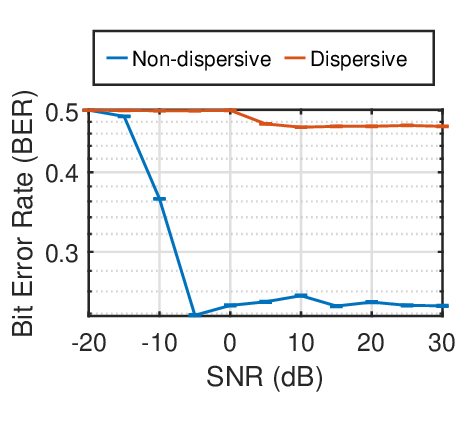}
  \caption{BER comparison between dispersive and non-dispersive channels across varying SNR levels. Results are averaged over 832 Monte Carlo trials. For the employed channel memory of \(T_{\mathrm{mem}}=10\) s, the omitted channel-energy fraction is \(1.15\times10^{-111}\).}
  \label{BER_SNR_Plot_disp_vs_nondisp}
\end{figure}

To compare dispersive and non-dispersive channels, the 4-symbol modulation scheme is simulated by taking \(\mu = 0.1\, m/s\), \(q_v=0.00524\,m/\sqrt{s}\) for different SNR values. In this configuration, the P\'eclet number was found as $4131.82$ by (\ref{eq:Peclet_number}), and the channel dispersion time was calculated as \(0.11 \, s\) by (\ref{eq:T_d_definition}). 1000 data symbols and 10 pilot symbols were transmitted with \(T_{sym} = 2 \, s\) corresponding to a non-dispersive channel and \(T_{sym} = 0.1 \, s\) corresponding to a dispersive channel. For a fair comparison, the simulation used fixed energy per bit. The resulting BER-SNR plot is given in \autoref{BER_SNR_Plot_disp_vs_nondisp}. The BER of the non-dispersive case decreases substantially as the SNR increases, although it approaches an error floor of approximately \(0.22\). In contrast, the BER of the dispersive case remains close to \(0.5\) throughout the investigated SNR range. Since increasing the SNR does not eliminate this performance gap, the observed degradation is predominantly attributed to channel dispersion rather than additive noise. The purpose of this comparison is to show that the derived dispersion time provides a principled criterion for assessing when the symbol duration should be increased because of channel effects.

\section{Orthogonality Loss Ratio}
In the context of this paper, an optimal pulse is defined as one that enables reliable communication at the highest rate allowed by the channel dynamics. The pulse design is governed by two key parameters: the symbol period \(T_{sym}\) and the signaling dimension \(N\). The OLR metric is specifically constructed to capture the distortion introduced by the channel on individual pulse waveforms. To isolate this effect, the analysis considers a single pulse transmission at a time and evaluates how the channel spreads the pulse energy into regions corresponding to other orthogonal pulse components for each pulse. In this sense, OLR quantifies the loss of orthogonality induced by channel dispersion. This metric is fundamentally different from BER, which reflects the overall system performance and depends on noise, symbol sequencing, and detection strategies. In contrast, OLR provides a structural measure of pulse separability after propagation, enabling a direct connection between channel-induced dispersion and the feasibility of pulse-based signaling. Since the purpose is to understand the effects of the channel on the pulse shape, the AWGN is ignored in this analysis.

\begin{figure}[!t]
  \centering

  \subfloat[]{
    \includegraphics[width=0.45\linewidth]{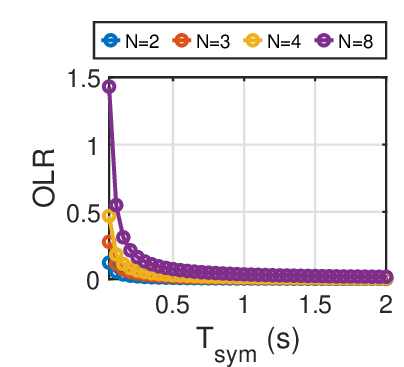}
    \label{fig:Pulse_Leakage_vs_Tsym_LTV_nondispersive}}
  \subfloat[]{
    \includegraphics[width=0.45\linewidth]{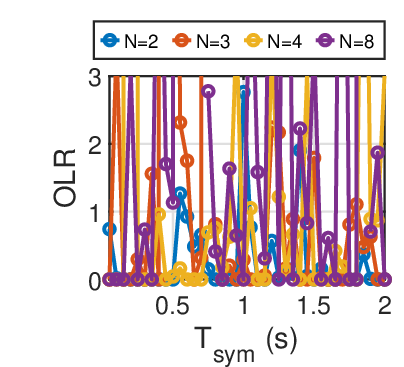}
    \label{fig:Pulse_Leakage_vs_Tsym_LTV_dispersive_y_limited}}

  \caption{Orthogonality Loss Ratio versus symbol duration \(T_{sym}\) for different signaling dimensions \(N\) for 
  \protect\subref{fig:Pulse_Leakage_vs_Tsym_LTV_nondispersive} advection-dominated flow and 
  \protect\subref{fig:Pulse_Leakage_vs_Tsym_LTV_dispersive_y_limited} diffusion-dominated flow. 
  }
  \label{Pulse_Leakage_vs_Tsym}
\end{figure}

Because the signals are nonnegative under particle communication, the only way to create orthogonal pulses is to separate them in the given symbol period so that only one of them is positive at a given time instance. Here, the symbol period is divided into \(N\) equal segments, each corresponding to one dimension in the signal space. For this analysis wind mean is taken as \(\mu = 0.5\, m/s\) and \(q_v = 3.16 \times 10^{-5} \,m/\sqrt{s}\) is set for an advection-dominated flow (\(P_e = 104434.7\)) and \(q_v = 0.4 \,m/\sqrt{s}\) is set for a diffusion-dominated flow (\(P_e = 4.42\)). To achieve reliable communication, after a pulse passes through the channel, the matched filters associated with the other orthogonal pulses should produce outputs that are as small as possible relative to the output of the filter associated with the transmitted pulse. Defining pulses as \(p_1, p_2, ..., p_N\) and using the nonnegativity of the signals, this observation can be formalized as
\begin{equation}
\label{eq:ideal_pulse_criterion}
\begin{split}
\text{OLR} = 
\frac{\sum_{i=1}^{N}
\sum_{\substack{j=1 \\ j \neq i}}^{N} c_i(t) * p_j^{\mathrm{MF}}(t)}
{\sum_{i=1}^{N} c_i(t) * p_i^{\mathrm{MF}}(t)},
\end{split}
\end{equation}
\noindent where \(c_i(t)\) is the channel output calculated using (\ref{eq:c_solution_point_source_phiD}) for an input pulse \(p_i(t)\). The numerator aggregates the outputs of the mismatched filters, and therefore represents the channel-induced leakage of each pulse into the other pulse dimensions. The denominator aggregates the corresponding matched-filter outputs and represents the desired response. Thus, the OLR measures undesired intra-symbol waveform interference relative to the desired response. A pulse set with a smaller OLR is better in terms of maintaining the separability of the transmitted waveforms after channel propagation, which directly improves the robustness of pulse-based signaling under diffusion–advection dynamics. The value of this sum is plotted against \(T_{sym}\) for different \(N\) values, where only a single pulse is transmitted to isolate the distortion introduced by the channel and receiver processing for flows dominated by advection and diffusion. The results are given in \autoref{Pulse_Leakage_vs_Tsym}. In advection-dominated flows, the rapid transport of particles preserves the temporal localization of pulses, resulting in reduced overlap between the channel-distorted responses of different pulses as \(T_{sym}\) increases. Consequently, although $N=2$ yields the lowest leakage for short symbol durations, larger pulse sets become viable at longer \(T_{sym}\), enabling higher-dimensional signaling. In contrast, for diffusion-dominated flows, increasing \(T_{sym}\) does not lead to a comparable reduction in OLR. This indicates that diffusion causes persistent temporal spreading that degrades pulse orthogonality regardless of symbol duration. As a result, the channel-induced mixing of pulses becomes intrinsic, rather than rate-dependent. While the previous analyses primarily focused on advection-dominated regimes where reliable symbol detection is achievable, diffusion-dominated channels represent a fundamentally different operating condition in which severe waveform mixing can prevent meaningful symbol-level decoding. In such cases, BER alone is insufficient to characterize system behavior, and structural metrics such as Orthogonality Loss Ratio OLR provide insight into the feasibility of pulse-based signaling itself. 

To examine whether the strong dispersion in \autoref{Pulse_Leakage_vs_Tsym}\protect\subref{fig:Pulse_Leakage_vs_Tsym_LTV_dispersive_y_limited} is associated with frequent backward transport permitted by the Gaussian wind model, we quantify the probability of net accumulated flow reversal along the transmitter-receiver direction. For the considered geometry, the projected accumulated flow displacement is defined as $A_{\mathrm{TR}}(\tau,t)\triangleq \frac{1}{\sqrt{2}}\int_{t-\tau}^{t}[v_1(\lambda)+v_2(\lambda)]\,d\lambda$. Here, \(A_{\mathrm{TR}}(\tau,t)<0\) represents a net accumulated flow reversal over the interval of length \(\tau\). Under the adopted white Gaussian wind model, \(A_{\mathrm{TR}}(\tau,t)\sim\mathcal{N}(\sqrt{2}\mu\tau,q_v^2\tau)\), and therefore \(\Pr[A_{\mathrm{TR}}(\tau,t)<0]=\Phi(-\sqrt{2}\mu\sqrt{\tau}/q_v)\), where \(\Phi(\cdot)\) is the standard normal cumulative distribution function. For the diffusion-dominated setting in \autoref{Pulse_Leakage_vs_Tsym}\protect\subref{fig:Pulse_Leakage_vs_Tsym_LTV_dispersive_y_limited}, the nominal advection time over the \(1~\mathrm{m}\) transmitter-receiver separation is \(1/(\sqrt{2}\mu)\approx1.41~\mathrm{s}\), for which the reversal probability is approximately \(1.77\%\). The corresponding probabilities at \(\tau=1~\mathrm{s}\) and \(\tau=2~\mathrm{s}\) are approximately \(3.85\%\) and \(0.62\%\), respectively. The probability decreases with the accumulation interval. These values show that the strong dispersion in \autoref{Pulse_Leakage_vs_Tsym}\protect\subref{fig:Pulse_Leakage_vs_Tsym_LTV_dispersive_y_limited} is not associated with net flow reversals.

\begin{figure}[!h]
  \centering
  \includegraphics[width=0.6\linewidth]{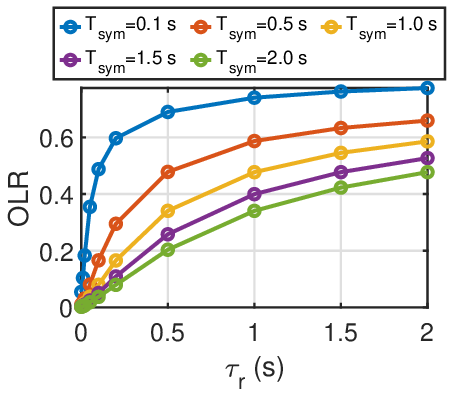}
  \caption{Orthogonality Loss Ratio versus receiver time constant $\tau_r$ for different symbol durations $T_{sym}$.}
  \label{OLR_vs_tau_r_FirstOrderReceiver}
\end{figure}

In addition to channel-induced distortion, practical receiver systems introduce additional temporal filtering effects due to finite sensor response and recovery times \cite{Correction_of_Dynamic_Errors_of_a_Gas_Sensor_Based_on_a_Parametric_Method_and_a_Neural_Network_Technique}. To study this effect, we adopt a simplified first-order receiver model based on the dynamic gas-sensor model presented in \cite{Correction_of_Dynamic_Errors_of_a_Gas_Sensor_Based_on_a_Parametric_Method_and_a_Neural_Network_Technique} and use the proposed OLR metric to analyze the combined effects of the channel and receiver dynamics on pulse orthogonality. The receiver is modeled as a first-order linear system governed by
\begin{equation}
\label{eq:first_order_Rx}
\begin{split}
\tau_r \frac{d\gamma_i(t)}{dt} + \gamma_i(t) = c_i(t),
\end{split}
\end{equation}
\noindent where $c_i(t)$ denotes the channel output corresponding to the transmitted pulse $p_i(t)$, $\gamma_i(t)$ is the continuous-time receiver output, and $\tau_r$ is the receiver time constant. The solution of this equation can be expressed as a convolution $\gamma_i(t) = (c_i * h_r)(t)$, where $h_r(t) = \frac{1}{\tau_r} e^{-t/\tau_r} u(t)$ is the impulse response associated with the first-order receiver dynamics. The overall system can be interpreted as a cascade of the diffusion–advection channel and the first-order receiver dynamics block, resulting in an effective impulse response that governs the observed waveform distortion. In this model, the receiver introduces additional temporal smoothing, which can further degrade the orthogonality of the pulse set beyond the distortion induced by the channel. Therefore, the OLR metric naturally extends to quantify the orthogonality loss of the overall linear system, including both channel and receiver effects. In the presence of receiver dynamics, the OLR in (\ref{eq:ideal_pulse_criterion}) is evaluated using the continuous-time receiver output $\gamma_i(t)$ instead of the channel output $c_i(t)$. To investigate the impact of receiver dynamics, OLR is evaluated as a function of the receiver time constant $\tau_r$ for different symbol durations $T_{sym}$. Commercial photoionization detector (PID) sensors can exhibit \(T_{90}\) response times below \(3~\mathrm{s}\), corresponding to \(\tau_r<3/\ln(10)\approx1.3~\mathrm{s}\) under the adopted first-order model \cite{MiniPID_2_Sensor_Guide}. A laser-absorption ammonia sensor reports \(10\%\)-\(90\%\) rise and \(90\%\)-\(10\%\) fall times of \(4.5~\mathrm{s}\) and \(3.7~\mathrm{s}\), respectively, corresponding to equivalent time constants of approximately \(2.05~\mathrm{s}\) and \(1.68~\mathrm{s}\) \cite{A_portable_sensor_for_in_situ_measurement_of_ammonia_based_on_near_infrared_laser_absorption_spectroscopy}. Accordingly, the OLR is evaluated over the representative range \(0\leq\tau_r\leq2~\mathrm{s}\), as presented in \autoref{OLR_vs_tau_r_FirstOrderReceiver}. The OLR increases monotonically with the receiver time constant $\tau_r$ for all considered symbol durations, indicating that slower receiver responses introduce additional temporal smoothing and consequently increase pulse mixing. This results in a progressive loss of orthogonality between the pulse components at the receiver. The effect is particularly pronounced for shorter symbol durations, where the transmitted waveforms vary rapidly in time and therefore cannot be accurately tracked by a receiver with limited temporal resolution. In contrast, for larger symbol durations, the pulse structure evolves more slowly, allowing the receiver to better follow the signal variations. These observations highlight that receiver dynamics play a critical role in determining pulse separability, and therefore must be jointly considered with channel dispersion when designing practical pulse-based communication schemes, especially in high-rate transmission scenarios where short symbol durations are employed.

In continuous transmissions, pulse overlap can degrade orthogonality. To characterize orthogonal-pulse separability in the presence of ISI, we extend the OLR definition and call the resulting metric the Interfered Orthogonality Loss Ratio (IOLR). For a length-\(K\) sequence, let \(i_k\in\{1,\ldots,N\}\) denote the index of the pulse \(p_{i_k}(t)\) transmitted in the \(k\)-th symbol interval. The resulting channel output \(c(t)\) contains the combined effects of all pulses in the sequence. The matched-filter outputs are sampled at \(t_k\), and the IOLR is defined as
\begin{equation}
\label{eq:interfered_OLR}
\mathrm{IOLR}
=
\frac{\sum_{k=1}^{K}\sum_{\scriptstyle j=1,\,j\neq i_k}^{N}
\left(c*p_j^{\mathrm{MF}}\right)(t_k)}
{\sum_{k=1}^{K}\left(c*p_{i_k}^{\mathrm{MF}}\right)(t_k)}.
\end{equation}
At each decision time \(t_k\), the numerator sums the outputs of the filters associated with pulses not transmitted in the \(k\)-th interval, while the denominator sums the output of the filter associated with the transmitted pulse \(p_{i_k}(t)\). Thus, the IOLR measures pulse separability after ISI has already been incorporated into \(c(t)\). The IOLR depends on the transmitted sequence because pulse ordering and repetition alter the combined channel output \(c(t)\). Since \(N^K\) different length-\(K\) sequences are possible, exhaustive evaluation becomes infeasible as \(N\) or \(K\) increases. The IOLR can instead be estimated using sufficiently long representative sequences selected according to the intended pulse usage. When multiple sequences are considered, the numerator and denominator terms in \eqref{eq:interfered_OLR} are separately accumulated over all sequences, and the final IOLR is calculated as the ratio of these accumulated quantities.

\section{Conclusion}

In this paper, a time-varying diffusion–advection channel is analytically characterized using the method of moving frames, yielding a closed-form, time-dependent channel impulse response. Based on this formulation, the channel is analyzed through its mean behavior and temporal correlation structure, enabling a principled characterization of channel memory. Channel dispersion time is defined as a physically meaningful time scale that governs symbol duration selection and achievable communication rates. This leads to a dispersion-based classification of diffusion–advection channels that directly links channel physics to communication system design. Building on this analytical foundation, an airborne particle communication system is investigated under directed wind conditions. The results demonstrate that, despite the fundamentally different propagation mechanisms, many core concepts from conventional digital communication remain applicable when appropriately adapted to mass-based signaling. Importantly, the analysis reveals that the communication space in airborne particle systems is not inherently binary. Instead, multi-symbol constellations can be realized using a single particle type by exploiting waveform design and receiver processing, shifting the modulation paradigm from molecule-limited to waveform-limited signaling. OLR  is introduced as a structural measure of pulse separability under channel-induced distortion. This metric provides insight into the feasibility of higher-dimensional signaling independent of symbol sequencing and noise, and reveals fundamental differences between advection-dominated and diffusion-dominated regimes. In particular, the results show that diffusion-dominated transport leads to intrinsic waveform mixing that limits reliable communication regardless of symbol duration.

\appendix[High-Péclet-Number Approximation of the Channel Dispersion Time]

Starting from the PDP in \eqref{eq:power_del_prof_expression}, we define its
moments as $M_n=\int_0^\infty \tau^nF_h(\tau)\,d\tau$ for $n=0,1,2$.
The mean delay is $\overline{\tau}=M_1/M_0$, and the RMS delay spread is
$T_d^2=M_2/M_0-(M_1/M_0)^2$. For white Gaussian wind, $D_{eff}=D+q_v^2$ is independent of $\tau$. Let $a=(x_3-z_{03})^2/(4D)$, $b=(x_3+z_{03})^2/(4D)$, and $B=\mu^2/D_{eff}$. Furthermore, define $A_1=L^2/(2D_{eff})+2a$,
$A_2=L^2/(2D_{eff})+a+b$, and $A_3=L^2/(2D_{eff})+2b$. We collect the $\tau$-independent terms in $C=\frac{D}{(4\pi D)^3D_{eff}} \exp\!\left(\frac{\mu[(x_1-z_{01})+(x_2-z_{02})]}{D_{eff}}\right)$. Expanding the exponent in \eqref{eq:power_del_prof_expression} and applying \cite[Eq.~(3.471.9)]{gradshteyn2007}, together with $K_{-\nu}(z)=K_\nu(z)$ \cite[\href{https://dlmf.nist.gov/10.27.E3}{Eq.~(10.27.3)}]{NIST_DLMF}, gives
\begin{equation}
\label{eq:rms_moment_closed_form}
\resizebox{0.95\linewidth}{!}{$
\begin{aligned}
M_n
=
2C
\sum_{j=1}^{3}
c_j
\left(
\frac{A_j}{B}
\right)^{(n-2)/2}
K_{\lvert n-2\rvert}
\left(
2\sqrt{A_jB}
\right),
\qquad n=0,1,2,
\end{aligned}
$}
\end{equation}
where the corresponding coefficients are $(c_1,c_2,c_3)=(1,-2,1)$. Denoting the $j=1$ contribution in \eqref{eq:rms_moment_closed_form} by $M_n^{(1)}$, the complete moment can be written as $M_n=M_n^{(1)}(1-R_{n,2}+R_{n,3})$, where
\begin{equation}
\label{eq:rms_image_term_ratios}
\begin{aligned}
R_{n,2}
&=
2
\left(
\frac{A_2}{A_1}
\right)^{(n-2)/2}
\frac{
K_{\lvert n-2\rvert}
\left(
2\sqrt{A_2B}
\right)
}{
K_{\lvert n-2\rvert}
\left(
2\sqrt{A_1B}
\right)
},
\\
R_{n,3}
&=
\left(
\frac{A_3}{A_1}
\right)^{(n-2)/2}
\frac{
K_{\lvert n-2\rvert}
\left(
2\sqrt{A_3B}
\right)
}{
K_{\lvert n-2\rvert}
\left(
2\sqrt{A_1B}
\right)
}.
\end{aligned}
\end{equation}
Let $z_j=2\sqrt{A_jB}$. From the integral representation of $K_\nu(z)$ \cite[\href{https://dlmf.nist.gov/10.32.E9}{Eq.~(10.32.9)}]{NIST_DLMF} and $\cosh t\geq1$, we have $K_\nu(z_j)/K_\nu(z_1)\leq e^{-(z_j-z_1)}$ for $j=2,3$. Moreover, $A_1\leq A_2\leq A_3$ and $(n-2)/2\leq0$ for $n=0,1,2$; hence, $R_{n,2}\leq2e^{-(z_2-z_1)}$ and $R_{n,3}\leq e^{-(z_3-z_1)}\leq e^{-(z_2-z_1)}$. Therefore,
\begin{equation}
\label{eq:rms_image_term_bound}
\resizebox{0.95\linewidth}{!}{$
\begin{aligned}
\left|
\frac{M_n}{M_n^{(1)}}-1
\right|
&=
\left|
-R_{n,2}+R_{n,3}
\right|
\leq
R_{n,2}+R_{n,3}
\leq
3e^{-(z_2-z_1)},
\end{aligned}
$}
\end{equation}
where $z_2-z_1 = P_e(\sqrt{1+ \frac{2D_{eff}x_3^2}{L^2D}}-1)$ for $x_3=z_{03}$. Allowing a relative moment error of at most $0.01$ requires $3e^{-(z_2-z_1)}\leq0.01$, or equivalently $z_2-z_1\geq\ln(300)\approx5.70$. Thus, this condition is sufficient for the combined magnitude of the $A_2$ and $A_3$ contributions to be at most $1\%$ of the $A_1$ contribution, permitting $M_n\simeq M_n^{(1)}$. Substitution of the dominant moment terms into the RMS delay-spread definition gives
\begin{equation}
\label{eq:rms_bessel_expression}
\begin{aligned}
T_d^2
\simeq
\frac{L^2}{2\mu^2}
\left[
\frac{
K_0(P_e)K_2(P_e)-K_1^2(P_e)
}{
K_2^2(P_e)
}
\right].
\end{aligned}
\end{equation}
Using the large-argument expansion in \cite[\href{https://dlmf.nist.gov/10.40.E2}{Eq.~(10.40.2)}]{NIST_DLMF}, with the coefficients given in \cite[\href{https://dlmf.nist.gov/10.17.E1}{Eq.~(10.17.1)}]{NIST_DLMF}, and defining $C_{P_e}=\sqrt{\pi/(2P_e)}e^{-P_e}$, gives
\begin{equation}
\label{eq:bessel_large_pe_expansions}
\begin{aligned}
K_0(P_e)
&=
C_{P_e}
\left[
1-\frac{1}{8P_e}
+O(P_e^{-2})
\right],
\\
K_1(P_e)
&=
C_{P_e}
\left[
1+\frac{3}{8P_e}
+O(P_e^{-2})
\right],
\\
K_2(P_e)
&=
C_{P_e}
\left[
1+\frac{15}{8P_e}
+O(P_e^{-2})
\right].
\end{aligned}
\end{equation}
When these expansions are substituted into \eqref{eq:rms_bessel_expression}, the common factor $C_{P_e}$ cancels. Expanding the denominators using $(1+u)^{-1}=1-u+O(u^2)$ then gives
\begin{equation}
\label{eq:bessel_ratio_large_pe}
\begin{aligned}
\left[
\frac{
K_0(P_e)K_2(P_e)-K_1^2(P_e)
}{
K_2^2(P_e)
}
\right]
=
\frac{1}{P_e}
+
O(P_e^{-2}).
\end{aligned}
\end{equation}
Consequently, $T_d^2\simeq L^2[1/P_e+O(P_e^{-2})]/(2\mu^2)$. Using $P_e=\sqrt{2}L\mu/D_{eff}$, the channel dispersion time satisfies
\begin{equation}
\label{eq:rms_delay_spread_high_pe}
\begin{aligned}
T_d
&\sim
\sqrt{
\frac{
LD_{eff}
}{
2\sqrt{2}\mu^3
}
},
\qquad
P_e\rightarrow\infty.
\end{aligned}
\end{equation}

\bibliographystyle{IEEEtran}
\bibliography{references.bib}

\vspace{11pt}

\begin{IEEEbiography}[{\includegraphics[width=1in,height=1.25in,clip,keepaspectratio]{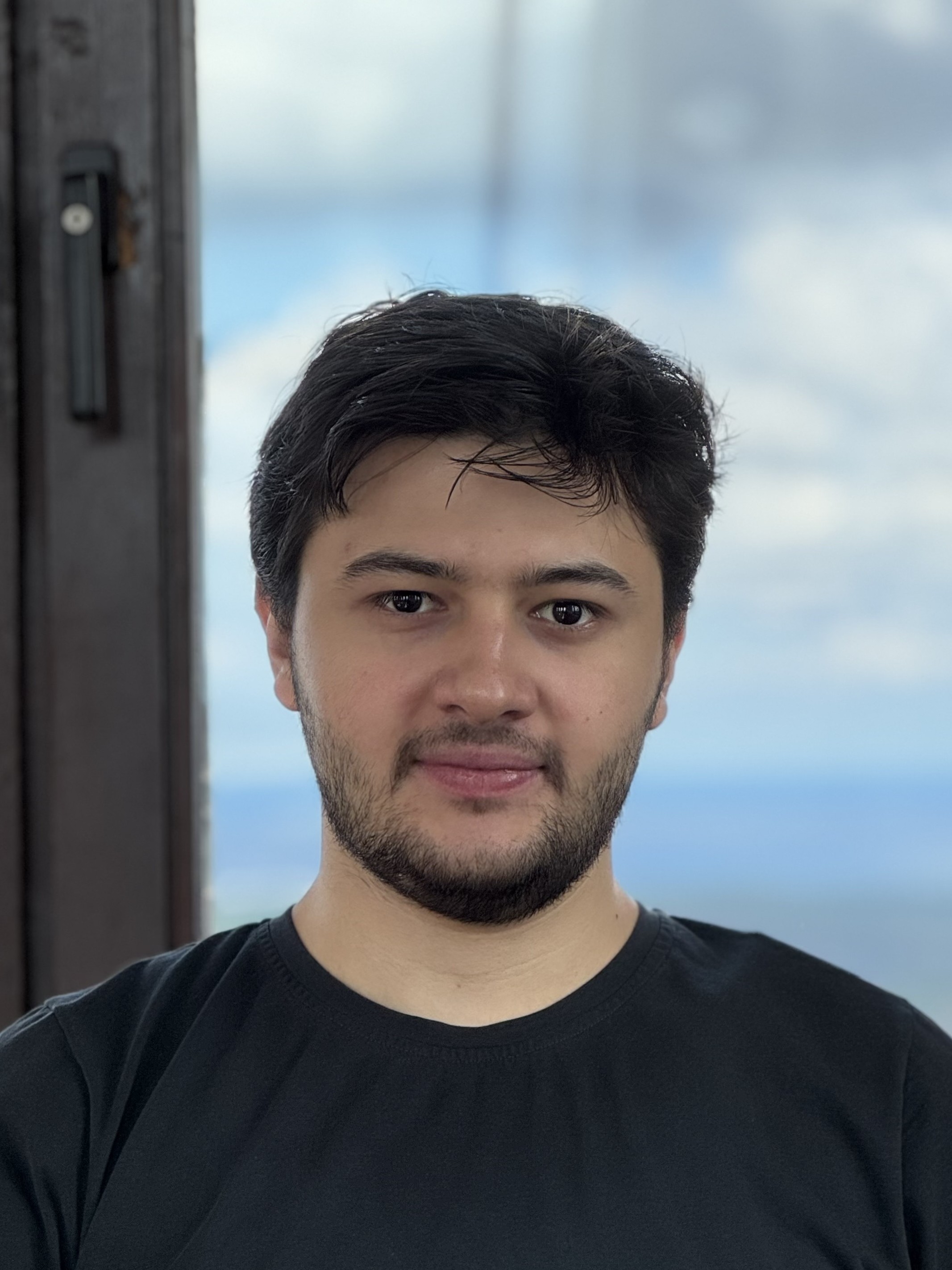}}]{Fatih Merdan}
completed his high school education at Kırıkkale Science High School. He received his B.Sc. degree in Electrical and Electronics Engineering from Middle East Technical University. He is currently pursuing his M.Sc. degree in Electrical and Electronics Engineering under the supervision of Prof. Akan at Koç University, Istanbul, Turkey.
\end{IEEEbiography}

\vspace{11pt}

\begin{IEEEbiography}[{\includegraphics[width=1in,height=1.25in,clip,keepaspectratio]{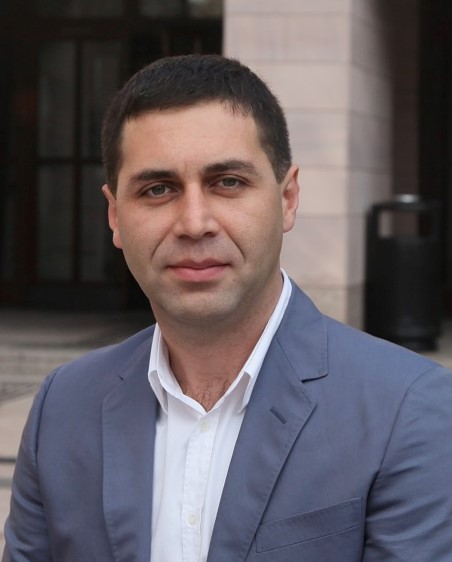}}]{Ozgur B. Akan}
\textbf{(Fellow, IEEE)} received the Ph.D. degree in electrical and computer engineering from Georgia Institute of Technology, USA, in 2004. He is currently a Professor in electronic and communication engineering with the University of Cambridge, U.K., where he leads the Centre for neXt Communications (CXC). He is also the Director of CXC, Koc¸ University, Turkiye. His research interests include wireless, space, quantum, and molecular communications.
\end{IEEEbiography}

\vfill

\end{document}